\documentclass[twocolumn]{aastex62}

\newcommand{\todo}{\ifmmode \text{\color{purple}\Huge{\(\bullet\)}} \else {\color{purple}{\Huge$\bullet$}}\fi}

\newcommand{\oone}{[O~{\sc{i}}]$\lambda6300$}
\newcommand{\otwo}{[O~{\sc{iii}}]$\lambda3727$}
\newcommand{\othree}{[O~{\sc{iii}}]$\lambda5007$}
\newcommand{\ntwo}{[N~{\sc{ii}}]$\lambda6584$}
\newcommand{\halpha}{H$\alpha$}
\newcommand{\hbeta}{H$\beta$}

\newcommand{\lbolothree}{L_{\rm bol,[OIII]obs}}
\newcommand{\lbolir}{L_{\rm bol,15 \mu m}}
\newcommand{\lothreeobs}{L_\mathrm{[OIII]obs}}
\newcommand{\lfifteen}{L_{\rm 15 \mu m}}
\newcommand{\eddington}{\lambda_\mathrm{Edd}}
\newcommand{\ledd}{L_{\rm Edd}}
\newcommand{\ltwelve}{L_{12 \mu{\rm m}}}
\newcommand{\mstar}{M_{\star}}
\newcommand{\msun}{M_{\odot}}
\newcommand{\mbh}{M_{\rm BH}}

\newcommand{\wise}{\textit{WISE}}

\usepackage{natbib}
\usepackage{amsmath}
\usepackage{multirow} 
\usepackage{textgreek}
\usepackage{xcolor}

\received{\today}
\submitjournal{ApJ}

\shorttitle{Luminosity declining AGN}
\shortauthors{Pflugradt et al.}

\begin{document}

\title{Finding of a population of active galactic nuclei showing a significant luminosity decline in the past $\sim 10^{3-4}$~yrs}

\correspondingauthor{Janek Pflugradt, Kohei Ichikawa}
\email{janek.pflugradt@astr.tohoku.ac.jp, k.ichikawa@astr.tohoku.ac.jp}

\author{Janek Pflugradt}

\affil{Astronomical Institute, Tohoku University, Aramaki, Aoba-ku, Sendai, Miyagi 980-8578, Japan}

\author[0000-0002-4377-903X]{Kohei Ichikawa}
\affil{Astronomical Institute, Tohoku University, Aramaki, Aoba-ku, Sendai, Miyagi 980-8578, Japan}
\affil{Frontier Research Institute for Interdisciplinary Sciences, Tohoku University, Sendai 980-8578, Japan}
\author[0000-0002-2651-1701]{Masayuki Akiyama}
\affil{Astronomical Institute, Tohoku University, Aramaki, Aoba-ku, Sendai, Miyagi 980-8578, Japan}
\author[0000-0001-6402-1415]{Mitsuru Kokubo}
\affil{Astronomical Institute, Tohoku University, Aramaki, Aoba-ku, Sendai, Miyagi 980-8578, Japan}
\affil{
Department of Astrophysical Sciences, Princeton University, Princeton, New Jersey 08544, USA
}
\author[0000-0003-2213-7983]{Bovornpractch Vijarnwannaluk}
\affil{Astronomical Institute, Tohoku University, Aramaki, Aoba-ku, Sendai, Miyagi 980-8578, Japan}
\author[0000-0001-6020-517X]{Hirofumi Noda}
\affiliation{Department of Earth and Space Science, Graduate School of Science, Osaka University, 1-1
Machikaneyama-cho, Toyonaka, Osaka 560-0043, Japan}
\author[0000-0003-2682-473X]{Xiaoyang Chen}
\affiliation{ALMA Project, National Astronomical Observatory of Japan, 2-21-1, Osawa, Mitaka, Tokyo 181-8588, Japan}



\begin{abstract}
Recent observations have revealed an interesting active galactic nuclei (AGN) subclass that shows strong activity at large scales ($\sim1$~kpc) but weaker at small scales ($<10$~pc), suggesting a strong change in the mass accretion rate of the central engine in the past $10^{3-4}$~yr.
We systematically search for such declining or fading AGN
by cross-matching the SDSS type-1 AGN catalog at $z<0.4$, covering the [O~{\sc{iii}}]$\lambda5007$ emission line which is a tracer for the narrow-line region (NLR) emission, with the \textit{WISE} mid-infrared (MIR) catalog covering the emissions from the dusty tori. Out of the 7,653 sources, we found 57 AGN whose bolometric luminosities estimated from the MIR band are at least one order of magnitude fainter than those estimated from the [O~{\sc{iii}}]$\lambda5007$ emission line. This luminosity declining AGN candidate population shows four important properties: 1) the past AGN activity estimated from the [O~{\sc{iii}}]$\lambda5007$ line reaches around the Eddington-limit, 2) more than 30\% of the luminosity declining AGN candidates show a large absolute variability of $\Delta W1 > 0.45$ mag in the previous $\sim10$~yr at
the \textit{WISE} 3.4~$\mu$m band,
3) the median ratio of $\log$([N~{\sc{ii}}]$\lambda6584/ \mathrm{H}\alpha\lambda6563)=-0.52$,
suggesting a lower gas metallicity and/or higher ionization parameter compared to other AGN populations.
4) the second epoch spectra of the population indicate a spectral type change for 15\% of the sources.
This population provides insights on the possible connection between the luminosity decline which started $\sim10^{3-4}$~yr ago and the decline in the recent $10$~yr.
\end{abstract}


\section{Introduction}\label{sec:intro}

One big question in the astronomy is on how super massive black holes (SMBH) increase their mass across the cosmic epoch in the universe.
Active galactic nuclei (AGN) are a key population for the SMBH growth since they are in a rapidly growing state of their BH masses through the gas accretion to the central SMBHs, until they reach the redshift-independent maximum mass limit at $M_\mathrm{max} \sim$ a few $\times 10^{10} \msun$ \citep{2003Netzer,2013Kormendy}.
 This indicates that SMBHs and their accretion systems might have a self-regulating process that shuts down the growth of SMBHs before reaching  
 a certain maximum mass ($\leq M_\mathrm{max}$)
 \citep{2009Natarajan,2016King,2016Inayoshi}.

One of the biggest unknowns for this accretion process is how long such an
AGN phase lasts.
Several studies indicate that 
the total lifetime of the AGN is $\sim 10^{7-9}$ yr \citep{1982Soltan, 2004Marconi} and
even the single episode has a length of $>10^5$ yr \citep{2015Schawinski} and likely around $10^{6-7}$~yr \citep{2004Marconi, 2006Hopkins},
which is still orders of magnitude longer than 
one person's lifetime of $\sim10^{2}$~yr.

One way to expand the AGN variability window
beyond the human lifetime
is to compare the activity of the different physical scales of the AGN, as the shutdown process propagates from the inner regions to the outer ones with the light crossing time  \citep[e.g., see ][]{2017Ichikawa_b}.
In the scale of 10 to 100 
gravitational radius ($R_\mathrm{g}$) away from the SMBH, 
the X-ray emitting corona and UV–optically bright 
accretion disk (AD) is located \citep{2010Dai,2010Morgan}, 
0.1-10 pc the mid-infrared (MIR) bright tori \citep{2013Burtscher}, 
and $\sim10^{3-4}$ pc the narrow line region \citep[NLR;][]{2002Bennert}.
Using those AGN components with different physical scales
enables us to explore the long-term luminosity variability with the order of $\sim10^{3-4}$~yr.
Recent studies have revealed 
that there are certain populations of AGN with strong activity in large scales but weaker ones in the small scales, which suggests a strong decline of the accretion
rate into the central SMBHs. These AGN are called fading or dying AGN.
Currently roughly $> 30$ of such fading AGN (candidates) were reported \citep{2013Schirmer,Kohei16,2017Kawamuro,2017Keel,2018Villar,2018Sartori,2018Wylezalek,Kohei19NeoRadio,Kohei19b,2019Chen,2020ChenSep,2020ChenDec,2020Esparza-Arredondo,2022Saade}.

In this paper, we conduct a systematic search of
such luminosity declining AGN by combining the SDSS AGN catalog with
the \textit{WISE} MIR all-sky survey. The SDSS AGN catalog
of \cite{MULLANEY} contains 25,670 AGN with the 
\othree~emission line, which is one of the large scale AGN indicators with $\sim$kpc scale. \textit{WISE} enables us to
obtain the warm dust emission heated by AGN, which
is one of the small physical scale AGN indicators with $\sim1$~pc scale.
The combination of these two AGN indicator information enables
us to obtain the long-term AGN variability of $\sim10^{3-4}$~yr timescale.
Throughout this paper, we 
adopt the same cosmological parameters 
as \cite{MULLANEY}; $H_0 = 71$~km~s$^{-1}$~Mpc$^{-1}$, $\Omega_\mathrm{M}=0.27$, and $\Omega_\Lambda=0.73$.

\section{Sample and Selection}\label{sec:seleintro}

\subsection{SDSS Type 1 AGN of \cite{MULLANEY}}
Our initial parent sample starts from the
AGN catalog compiled by \cite{MULLANEY}, which includes
25,670 optically selected active galactic nuclei (AGN)
from the SDSS DR7 data release \citep{SDSSDR7}
at $z<0.4$, where \halpha~is within the SDSS wavelength coverage.
\cite{MULLANEY} conducted their own spectral fitting
of the SDSS spectra including the narrow components of the narrow H$\beta$
and H$\alpha$ emission lines, 
which gives better fitting results for the emission lines
compared to the original
SDSS fitting method which applied a more simplified fitting \citep[e.g.,][]{yor00,SDSSDR7}.

In this study, we used only type-1 AGN from the \cite{MULLANEY} catalog for searching luminosity declining AGN because of the two reasons; 1) the more reliable observed \othree~luminosities with small extinctions considering their viewing angle \citep[e.g.,][]{ant93} and 2) availability of $\mbh$ measurements thanks to the existence of the broad
emission lines.
\cite{MULLANEY} classified AGN as type-1 if they fulfill the following criteria for their \halpha-line:
1) an extra Gaussian (for the broad component) in addition to the narrow Gaussian provides a significantly better fit,
2) the broad component flux exceeds the narrow one, and 3)
FWHM of broad \halpha~component has $>600$~km~s$^{-1}$.

The obtained type-1 AGN catalog contains 9,455 sources and it provides the emission line fluxes, redshifts, and luminosities of \othree, \halpha, \hbeta, and \ntwo,
and the obtained fluxes are separated into narrow and broad components\footnote{We applied the $(1+z)$ $k$-correction for the \textit{B}-band continuum and emission line fluxes including H$\alpha$ luminosities since \cite{MULLANEY} did not apply for this correction, and that also changes the black hole mass estimation as well by a factor of $(1+z)^{0.31}$. We also confirmed that this $(1+z)$ correction does not change our main results since the median redshift of our sample is $\left<z\right>=0.2$, which is corresponding to a factor of 1.2 for luminosities and a factor of 1.06 for the black hole mass estimations.},
as well as the physical values based on those emission line measurements, such as black hole masses ($\mbh$), and Eddington ratio ($\eddington$), both of which are estimated based on the broad H$\alpha$ emission lines \citep{HalpaSMBHmass}.

We then limited our sample to the sources with
reliable spectral fitting.
First, we limit our samples to low extinction values for the narrow-line region with $A_\mathrm{V}<10$, since otherwise the luminosities of these sources are unrealistically over-estimated \citep[see][ for more details]{MULLANEY}.
This reduces the sample into 9,372 sources.
Second, 
we limit the sources whose 
\othree~emission line is significantly detected, with
a signal to noise ratio of $\mathrm{S/N} \geq 5$.
The resulting type-1 AGN sample contains 7,755 sources.

It should be pointed out that searching only for type-1 luminosity declining AGN
 significantly reduces the possibility for finding genuine ``dying'' AGN
 whose central engine is completely quenched, such as Arp~187 \citep{Kohei19NeoRadio,Kohei19b}, which is likely
classified as type-2 AGN in the optical spectra because of the lack of the broad emission lines already in the current SDSS spectra.
We will explore luminosity declining/dying type-2 AGN systematic search in the forthcoming paper.

\subsection{Cross-matching with \textit{WISE}}

The \wise~mission mapped the entire sky in 3.4 $\mu$m (\textit{W1}), 4.6 $\mu$m (\textit{W2}), 12 $\mu$m (\textit{W3}), and 22 $\mu$m (\textit{W4}) bands \citep{WISE_WRIGHT10,WISE_DOI}.
In this study, we obtained the data from the latest \textit{ALLWISE}
catalog \citep{2013Cutri}.
We used the pipeline measured magnitudes at the \textit{W3}-band
based on the PSF-profile fitting on $\sim6$~arcsec scale,
called profile fitting magnitude and \texttt{w3mpro}
 in the \wise~catalog terminology.
The positional accuracy based on cross-matching
with the 2MASS catalog is $\sim2$~arcsec at the $3\sigma$ level
\citep[e.g., see][]{ich12,ich17a}, and we also applied the
2~arcsec cross-matching radius between the SDSS 
optical coordinates and \wise. 
This reduces the sample from 7,755 to 7,723 sources.

We obtained the \textit{W3}-band fluxes and treated the value as detection for the sources with \texttt{ph\_qual=A,B,C}, with SN higher than 2.
The \texttt{ph\_qual=U} were treated as upper limit detection since we are looking for low luminosity AGN in the MIR band and it is therefore also important to consider the weak detections in the MIR bands for our parent sample. We also applied the contamination-free sources
with \texttt{ccflag=0}. This reduces the sample to 7,653, where 6,028/1,496/75/54 of the sources have the photometric quality of \texttt{ph\_qual=A/B/C/U}, respectively.
We refer to this sample as ``parent sample''
for searching the luminosity declining AGN.

\subsection{AGN Luminosities and Bolometric Corrections}
\label{chap:bolo}
\begin{figure*}
    \begin{center}
    \includegraphics[width=0.7\textwidth]{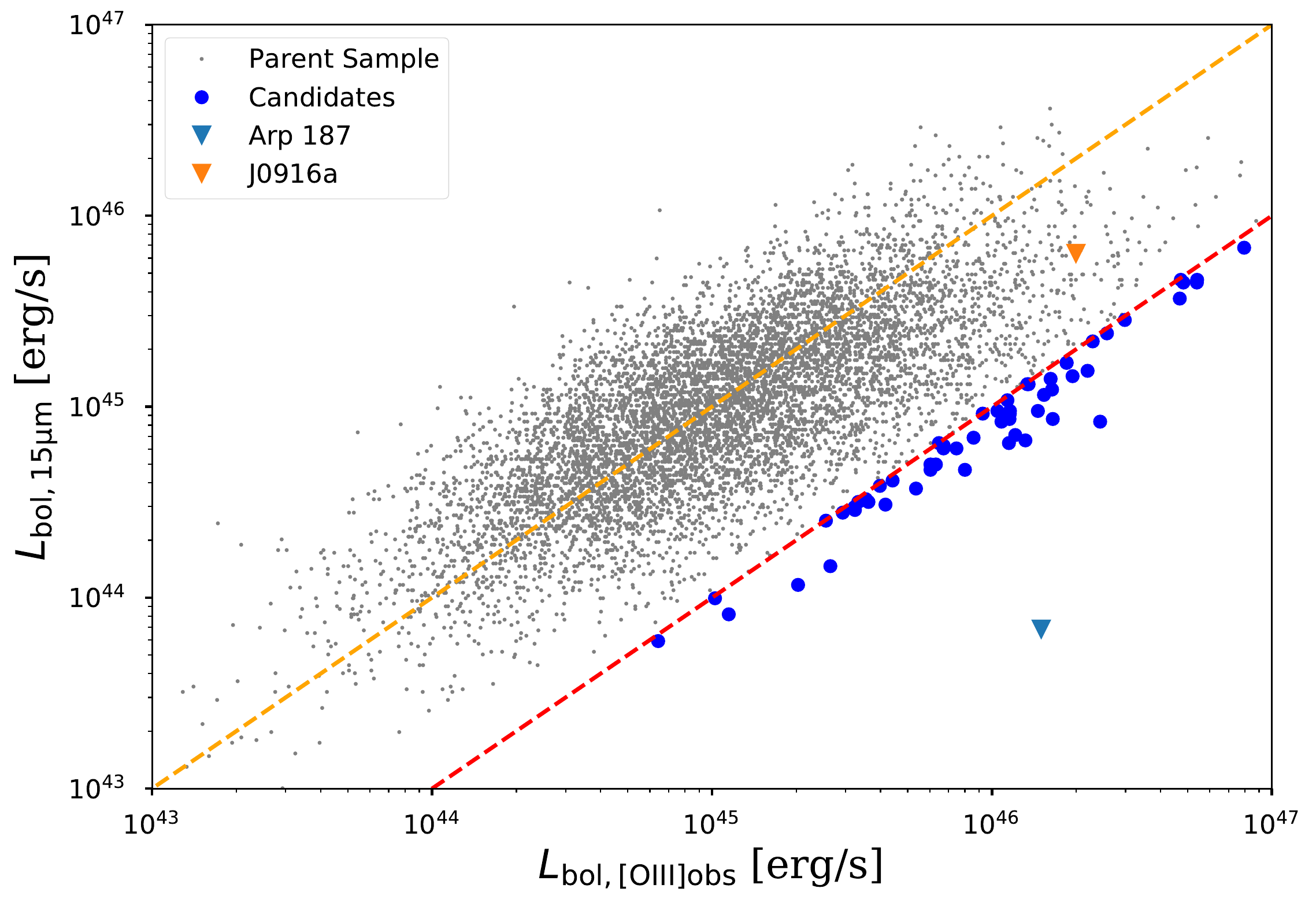}~
    \caption{
    The correlation of the bolometric AGN luminosity estimated from the \textit{WISE} 12~$\mu$m band and \othree~ emission line.
    The gray dot represents the parent sample and the blue filled circle represents the luminosity declining AGN candidates.
    The orange dashed line represents the 1:1 line of the two luminosities.
    For the comparison with other fading/dying AGN, we showed the location of Arp 187 (blue uppe-bound), which is one of the most promising dying AGN \citep{Kohei16,Torus_size_12um_Ichikawa_2017,Kohei19NeoRadio,Kohei19b} and 
    the location of J0916a (orange upper-bound), which is a fading AGN whose fading phase is likely recently started ($<10^2$~yr) with strong past outburst activity in the NLR \citep{2019Chen,2020ChenSep,2020ChenDec}.
    The selection criteria of the luminosity declining AGN candidates discussed in section \ref{Chap:SelectionCandi} is represented by the red dashed line.}
    \label{fig:boloplot}
    \end{center}
\end{figure*}

We measured the AGN bolometric luminosities based on the obtained AGN indicators tracing different physical scales.
For the NLR luminosities tracing a $\sim10^{3-4}$~pc scale, we utilized the 
observed \othree~luminosities obtained by \cite{MULLANEY}
and then applied the constant bolometric correction of
$\lbolothree = 3500 \lothreeobs$ with the median error
of 0.38~dex \citep[e.g.,][]{OIIILUMCorrecktion_Heckman_2004}.
We note that our sample contains [OIII] luminous AGN that are not covered in \cite{OIIILUMCorrecktion_Heckman_2004},
notably the sources with $\lothreeobs \geq 10^{42.5}$~erg~s$^{-1}$. 
We here apply the same constant bolometric correction even 
for those sources since the overall spectral shape does not change in the standard disk regime even when the accretion rate changes \citep{2008Kato}.
One might also wonder that the bolometric correction
might change in the super-Eddington regime.
We will discuss this point later in Section~\ref{sec:AGNlightcurve}.
Another point is that \cite{MULLANEY} already provide the bolometric luminosities of the sample by utilizing the extinction corrected \othree~luminosities. 
However, we did not use those since
they are often unrealistically large values
exceeding $\lbolothree>10^{48}$~erg~s$^{-1}$, which is the Eddington limit of the mass of $M_\mathrm{BH}\sim 10^{10}\msun$;
the known maximum mass limit of the SMBHs.

For the AGN dust luminosities tracing $\sim10$~pc scale,
we first derived the rest-frame 15~$\mu$m luminosities
by applying the $k$-correction from the obtained \textit{WISE} 12~$\mu$m flux density.
The rest-frame 15~$\mu$m flux density was  extrapolated from the obtained observed 12~$\mu$m flux density with the assumption of AGN IR spectral template of \cite{mul11} and with the obtained redshift.
The AGN bolometric luminosities from the AGN dust were finally estimated from the bolometric correction curve by \cite{Lum_functions_Hopkins07}, which has a typical bolometric correction value of $\lbolir \sim 10 \lfifteen$ with a scatter of a factor of 2.

In addition, we also measured the bolometric AGN luminosities from the accretion disk
emission by using the rest-frame \textit{B}-band continuum fluxes, tracing $\sim100 R_\mathrm{g}$ scale assuming the standard thin-disk accretion \citep[e.g.,][]{2008Kato}. The \textit{B}-band fluxes were calculated from the continuum of the SDSS spectra by \cite{MULLANEY}, and assumed that the continuum is dominated by the AGN accretion disk. For the bolometric correction, we applied the one by \cite{Lum_functions_Hopkins07}.

In summary, we obtained the three AGN bolometric luminosities from the NLR, AGN dusty torus, and the accretion disk.
Figure~\ref{fig:boloplot} shows
bolometric luminosity correlation between the one from NLR and the one from AGN dusty torus, showing a rough 1:1 relation each other.

\subsection{The Emission Region Size of Each AGN Indicator}
\label{chap:Sizecal}
We estimate the emitting size of each AGN component
by utilizing either theoretical or empirical luminosity-distance relations.
The emitting size from the SMBH to the \textit{B}-band
in the optical can be obtained by assuming the
standard geometrically thin $\alpha$-disk \citep[e.g.,][]{2008Kato}, and it is expressed as
\begin{align}
\begin{split}
D_\mathrm{AD}/\mathrm{AU}=886.3 \left(\dfrac{\lambda}{4450 \mathrm{\AA}}\right)^{4/3}&\left(\dfrac{M_\mathrm{BH}}{10^8 \msun}\right)^{1/3} \\& \left(\dfrac{L_{\lambda,\mathrm{bol}}}
{10^{46}\mathrm{ ~erg~s}^{-1}}\right)^{1/3},
\end{split}
\end{align}
here we assume the radiation efficiency $\eta_\mathrm{rad}=0.1$  \citep{1982Soltan} and $\lambda=4450\mathrm{\AA}$ (\textit{B}-band) as a tracer of the accretion disk \citep{1982Malkan}.

The size of the MIR dust emission region heated by AGN is obtained by MIR high spatial resolution  interferomerty observations 
\citep[e.g.,][]{2011Kishimoto}
and can be written as
\begin{align}
    D_\mathrm{torus}/\mathrm{pc}=  1.3\left(\dfrac{L_{\rm bol,15 \mu m}}{ 10^{46}\mathrm{~ erg ~s}^{-1}}\right)^{0.01}.
\end{align}


The size of the NLR and their AGN luminosity dependence has been studied by several authors \citep{2002Bennert,2013Hainline,2014Husemann}. Here we apply the relation between the NLR size and the AGN luminosity by \cite{NLR_OIII_Bae_2017} because they utilized the observed \othree~ luminosities as a tracer of AGN luminosity, which is  suitable for our sample of type-1 AGN. The equation is given as
\begin{align}
    D_\mathrm{NLR}/\mathrm{kpc}=2.4\left(\dfrac{L_{\rm bol,[OIII]obs}}{ 10^{46}\mathrm{~ erg ~s}^{-1}}\right)^{0.41}
\end{align}
The median size of the NLR of our parent sample is 
$D_\mathrm{NLR}\sim1$~kpc, which is consistent with our assumption that
NLR is a tracer of the past AGN activity of $10^{3-4}$~yr.

\subsection{Selection of luminosity declining AGN Candidates}
\begin{deluxetable*}{clll}[h!]
\tabletypesize{\footnotesize}
\tablecolumns{4}
\tablewidth{20pt}
\tablecaption{Column descriptions for the table of the selected luminosity declining AGN candidates. \label{tab:canditable_MR}}
\tablehead{
\colhead{Column}
 & \colhead{Header Name} & \colhead{Unit} & \colhead{Description}}
\startdata
1  & OBJID                &  & SDSS DR7 Object ID \\
2  & z                    &  & Spectroscopic Redshift\\
3  & ra                   & ° & Right ascension\\
4  & dec                  & °&  Declination\\
5  & SMBH\_MASS           & $10^7\msun$ & $\mbh$\\
6  & SMBH\_MASS\_ERR      & $10^7\msun$ & $\Delta\mbh$\\
7 & F\_12um           & $10^{-17}$ erg s$^{-1}$ cm$^{-2}$ &  $F_{\mathrm{12}\mu\mathrm{m}}$              \\
8 & F\_12um\_ERR      & $10^{-17}$ erg s$^{-1}$ cm$^{-2}$ & $\Delta F_{\mathrm{12}\mu\mathrm{m}}$                \\
9 & L\_12um            & erg s$^{-1}$ &   $\ltwelve$             \\
10 & L\_12um\_ERR       & erg s$^{-1}$ &  $\Delta\ltwelve$               \\
11  & L\_15um\_BOLO      & erg s$^{-1}$ & $\lbolir$ estimated for the torus               \\
12 & L\_15um\_BOLO\_ERR & erg s$^{-1}$ & $\Delta \lbolir$ estimated for the torus\\
13&PH\_QUAL\_12um && Photometric quality for the 12$\mu$m-band (A:SN$>$10 \& 1BSN$>$3 \& C:SN$>$2)\\
14 & F\_NII            & $10^{-17}$ erg s$^{-1}$ cm$^{-2}$ & $F_{\mathrm{[NII]}}$               \\
15 & F\_NII\_ERR       & $10^{-17}$ erg s$^{-1}$ cm$^{-2}$ &  $\Delta F_{\mathrm{[NII]}}$              \\
16 & L\_NII             & erg s$^{-1}$ &  $L_{\mathrm{[NII]}}$              \\
17 & L\_NII\_ERR        & erg s$^{-1}$ &  $\Delta L_{\mathrm{[NII]}}$      \\
18 & F\_HA             & $10^{-17}$ erg s$^{-1}$ cm$^{-2}$ &   $F_{\mathrm{H}\alpha}$              \\
19 & F\_HA\_ERR        & $10^{-17}$ erg s$^{-1}$ cm$^{-2}$ &  $\Delta F_{\mathrm{H}\alpha}$                \\
20 & L\_HA              & erg s$^{-1}$ &   $L_{\mathrm{H}\alpha}$             \\
21 & L\_HA\_ERR         & erg s$^{-1}$ &    $\Delta L_{\mathrm{H}\alpha}$             \\
22 & F\_OIII           & $10^{-17}$ erg s$^{-1}$ cm$^{-2}$ &  $F_\mathrm{[OIII]obs}$             \\
23 & F\_OIII\_ERR      & $10^{-17}$ erg s$^{-1}$ cm$^{-2}$ &   $\Delta F_\mathrm{[OIII]obs}$             \\
24 & L\_OIII            & erg s$^{-1}$ & $\lothreeobs $              \\
25 & L\_OIII\_ERR       & erg s$^{-1}$ &  $\Delta\lothreeobs$\\
26 & L\_OIII\_BOLO      & erg s$^{-1}$ & $\lbolothree$ estimated for the NLR\\
27 & L\_OIII\_BOLO\_ERR & erg s$^{-1}$ & $\Delta\lbolothree$ estimated for the NLR\\
28 & F\_HB             & $10^{-17}$ erg s$^{-1}$ cm$^{-2}$ &    $F_{\mathrm{H}\beta}$              \\
29 & F\_HB\_ERR        & $10^{-17}$ erg s$^{-1}$ cm$^{-2}$ &  $\Delta F_{\mathrm{H}\beta}$              \\
30 & L\_HB              & erg s$^{-1}$ &  $L_{\mathrm{H}\beta}$               \\
31 & L\_HB\_ERR         & erg s$^{-1}$ &    $\Delta L_{\mathrm{H}\beta}$             \\
32 & F\_OPTICAL        & $10^{-17}$ erg s$^{-1}$ cm$^{-2}$ &  $F_{\mathrm{\textit{B}-band}}$              \\
33 & L\_OPTICAL         & erg s$^{-1}$ & $L_\mathrm{\textit{B}-band}$               \\
34 & L\_OPTICAL\_BOLO   & erg s$^{-1}$ & $L_{\rm bol, \textit{B}-band}$ estimated for the accretion disk               \\
35 & SIZE\_NLR        & pc &  $D_\mathrm{NLR}$               \\
36 & SIZE\_TORUS           & pc & $D_\mathrm{torus}$                \\
37 & SIZE\_AD      & pc &  $D_\mathrm{AD}$              \\
38 & EDD\_NLR        &  &  $\eddington$ estimated for the NLR out of the \othree~line information             \\
39 & EDD\_TORUS           &  &  $\eddington$ estimated for the torus out of the MIR photometric information               \\
40 & EDD\_AD      &  &  $\eddington$ estimated for the accretion disk out of the optical photometric information\\
41&LOG\_R& & $\log(R)$\\
42&DELTA\_W1 & mag & $\Delta W1$\\
43&MJD\_WISE& & MJD of the \wise~observation\\
44&MJD\-SDSS& & MJD of the SDSS observation
 \enddata  
  \thispagestyle{empty}
\tablecomments{Table \ref{tab:canditable_MR} is published in its entirety in the machine-readable format. All column names, their units, and descriptions are shown here for guidance regarding its form and content.}
\end{deluxetable*}

\label{Chap:SelectionCandi}
\begin{figure}
    \includegraphics[width=0.47\textwidth]{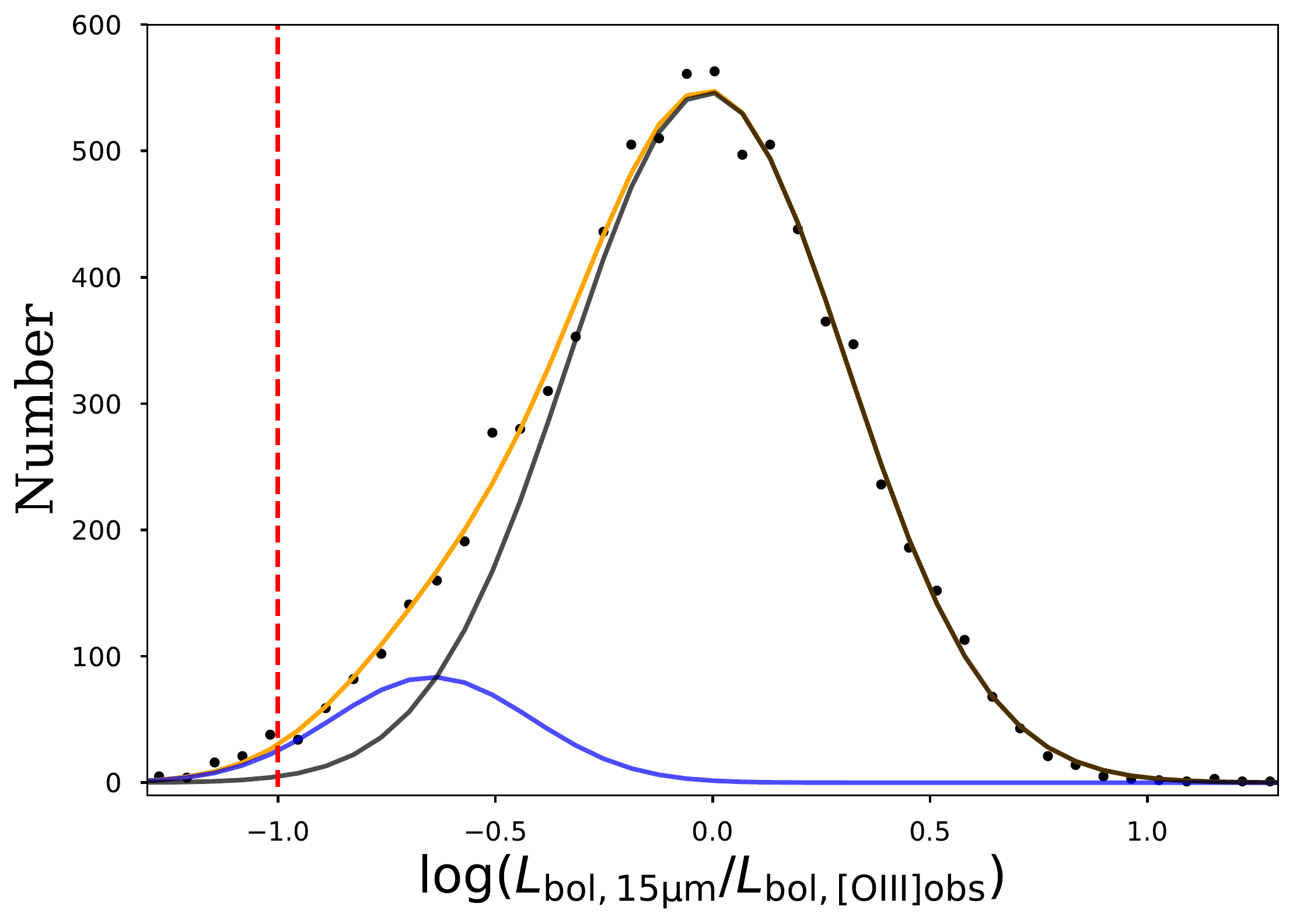}
    \caption{The binned distribution of $\log(R)$ of the parent sample is represented here by the black dotes. The black and blue Gaussian line represents the ones as a result of the two Gaussian component fitting. 
    The orange line represents the composite of the two components. The red line shows the selection cut for the luminosity declining AGN candidates at $\log(R)=-1$.}
    \label{fig:ratioselection}
\end{figure}
In order to search for luminosity declining AGN candidates, it is necessary to compare two AGN indicators, a large-scale indicator with a short-scale indicator.
NLR is a promising tool as a large-scale AGN indicator because of its large size with $\sim1-10$ kpc, and the \othree~emission line
is one of the good indicators of NLR luminosity \citep{2002Bennert}.
On the other hand, the emission from the dusty tori is a good AGN indicator tracing smaller physical scale of $\sim10$~pc \citep{2013Burtscher,2005Packham,2008Radomski,2009Ramos,2011Alonso,2015Ichikawa,2004Jaffe,2009Raben,2012Honig,2013Honig,2013Burtscher,2014Tristram,2016Lopez,lop18} and 
can be traced by the MIR luminosities \citep[e.g.,][]{2009Gandhi,ich12,ich17a,ich19,nik21a,nik21b}. 

Figure~\ref{fig:boloplot} shows 
the luminosity correlation between
the two AGN bolometric luminosities from different AGN indicators,
NLR ($\sim$kpc) and dusty tori ($\sim10$~pc), considered to be tracing
the past AGN activities of $\sim3000$~yr and $\sim30$~yr ago \citep{Torus_size_12um_Ichikawa_2017}. Figure~\ref{fig:boloplot} shows a
one-by-one relation for most of the sample, while
some sources have a significantly smaller value at $\lbolir/\lbolothree < 0.1$, showing with blue points below the red dashed line.

Figure~\ref{fig:ratioselection} shows the
histogram of the logarithmic ratio $\log(R)$;
defined as 
\begin{align}
    \log(R) = \log \left( \frac{\lbolir}{\lbolothree}\right).
\end{align}
Here we used the Scott function \citep{sco79} to define the optimal bin width.
We conducted one Gaussian fitting of the histogram of $\log(R)$ 
and all three parameters 
(amplitude, mean and standard deviation)
were set as free.
The fitting still leaves an excess of the distribution at $\log(R) \sim -1$, indicating an additional component, which is also suggested from the blue point sources in Figure~\ref{fig:boloplot}.
We then applied the two Gaussian fitting with all parameters set as free,
and the second Gaussian component nicely reproduces the second peak at  $\log(R) \sim -0.7$.
We also checked the reduced $\chi^2$ values
of the fittings,
with {$\chi^2_\nu=3.5$} for two Gaussians and $\chi^2_\nu=5.4$ for one Gaussian fitting.
This indicates that the distribution
can be well described with at least 
two Gaussian components as shown in Figure~\ref{fig:ratioselection}, and it also suggests that there is a significant peak at $\log(R) \sim -0.7$,
which might be related to a population of luminosity declining AGN candidates in this study.

Based on the histogram shown in Figure~\ref{fig:ratioselection},
we selected AGN with $\log(R)<-1$ and we refer to them as 
``luminosity declining AGN candidates'' hereafter.
As shown in Figure~\ref{fig:boloplot}, our selection cut
is much more conservative considering the location of another reported fading AGN SDSS J0916a in the plane, which has likely started luminosity decline in the last $<10^2$~yr \citep{2020ChenDec}.
On the other hand, most of our sample spans with $-1.5<\log(R) < -1$, which is
one order of magnitude larger value than that of Arp~187, which is a dying AGN,
whose central engine is completely quenched. This is also a natural outcome
considering that our sample is type-1 AGN whose broad line region still exists,
even if they might be in a luminosity declining phase.

As a result, 57 luminosity declining AGN candidates were selected,
which is $\sim$0.7\% of the parent sample. 
Table \ref{tab:canditable_MR} summarizes the list of the physical parameters of the luminosity declining AGN candidates in this study.

\section{Results}

\subsection{Basic Sample Properties}

We here summarize the BH
properties of the obtained 57 luminosity declining AGN candidates.
First, we show the basic differences between
the luminosity declining AGN candidates and the parent sample.
Then, we show the properties of luminosity declining AGN candidates
on the SMBH activities.

\subsubsection{AGN Luminosity and Redshift Plane}\label{sec:LAGN_vs_z}

\begin{figure*}
\begin{center}
\includegraphics[width=0.48\textwidth]{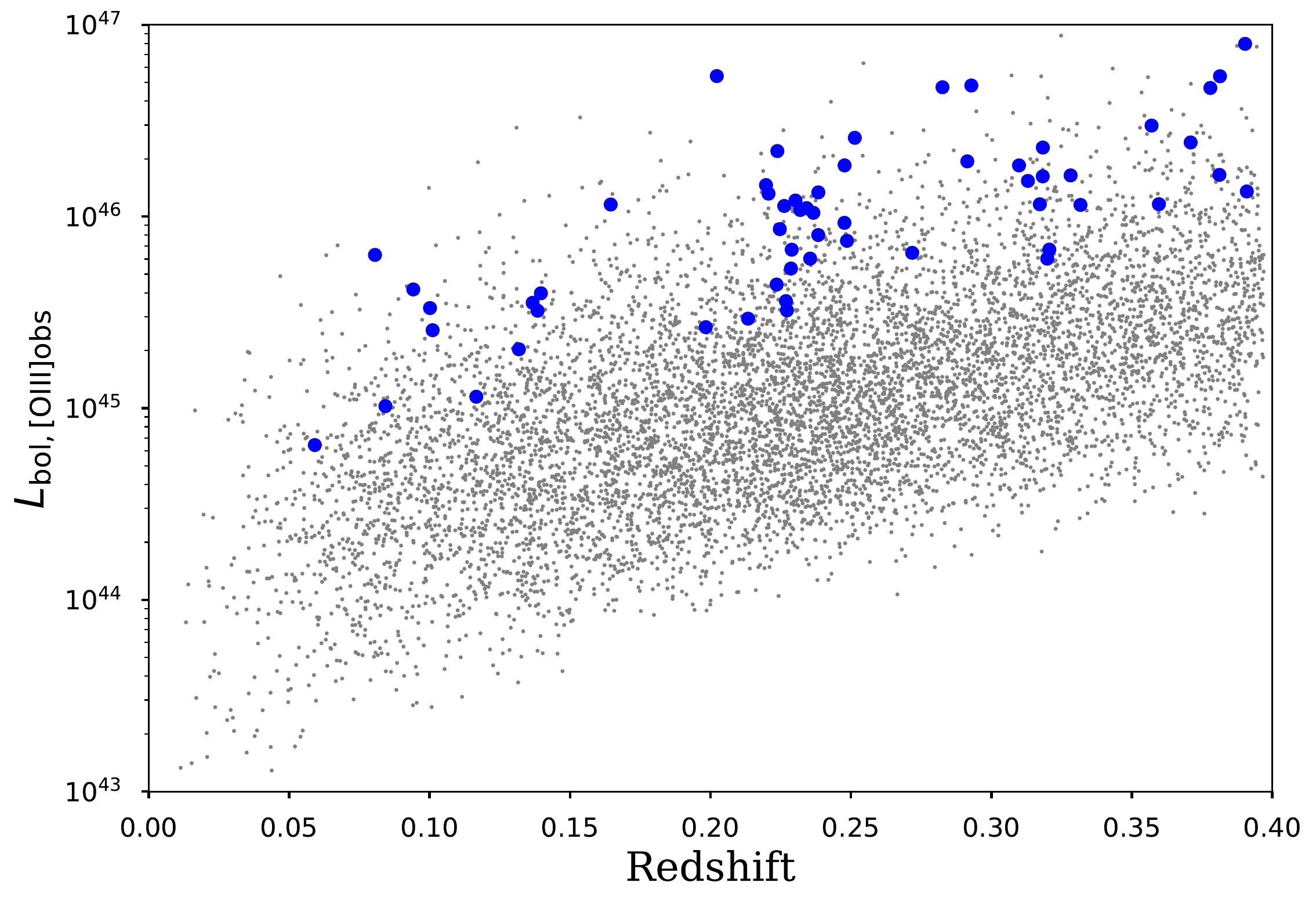}~
\includegraphics[width=0.48\textwidth]{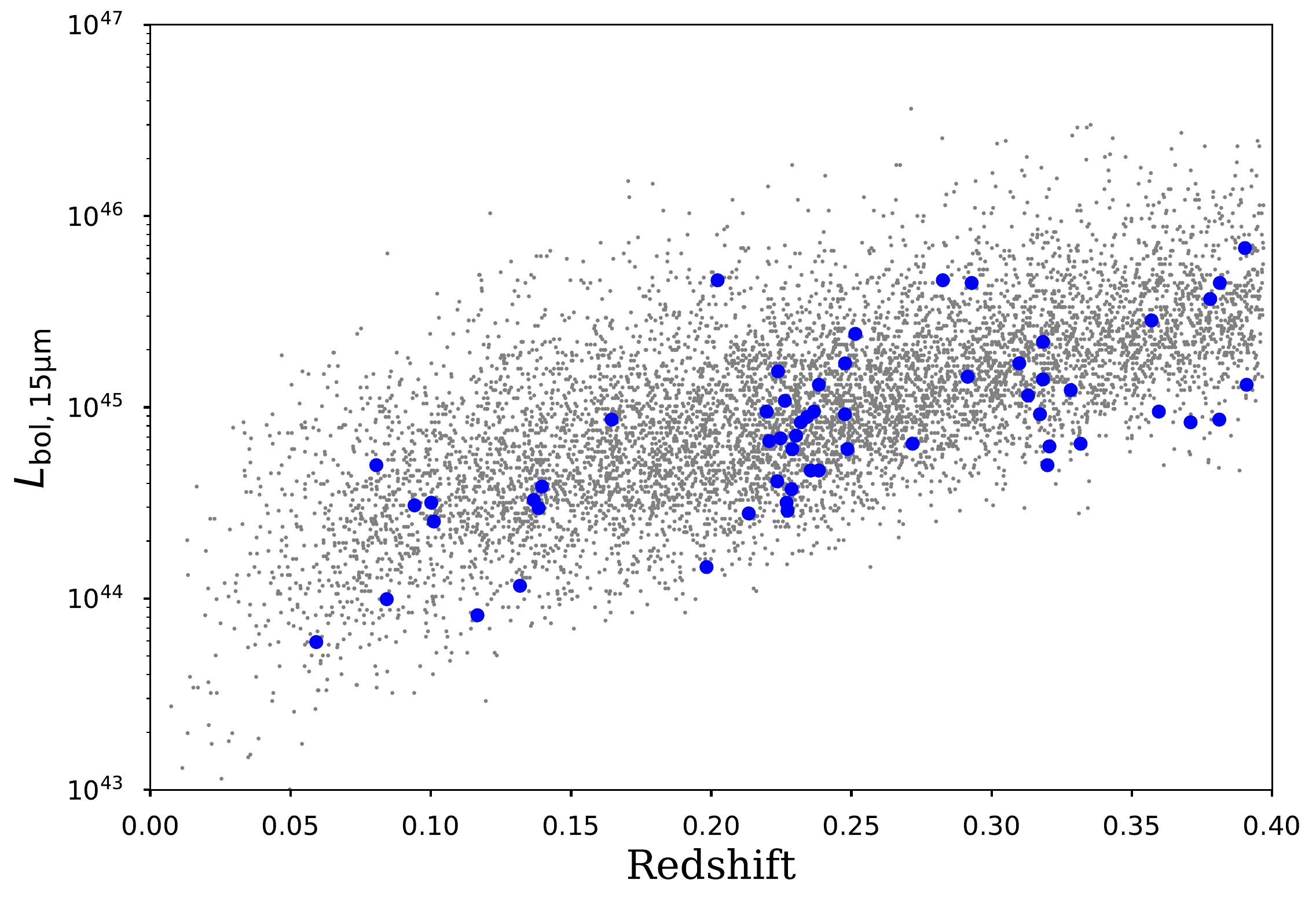}~
\caption{
Luminosities of luminosity declining AGN candidates (blue filled circle) and the parent sample
(gray dot) as a function of redshift.
(Left) The bolometric luminosity estimated from the observed \othree~emission luminosities ($\lbolothree$) as a function of redshift. (right)
The bolometric luminosity estimated from the 15~$\mu$m bands ($\lbolir$) as a function of redshift.
}\label{fig:Redshift}
\end{center}
\end{figure*}

\begin{figure}
    \centering
    \includegraphics[width=1\linewidth]{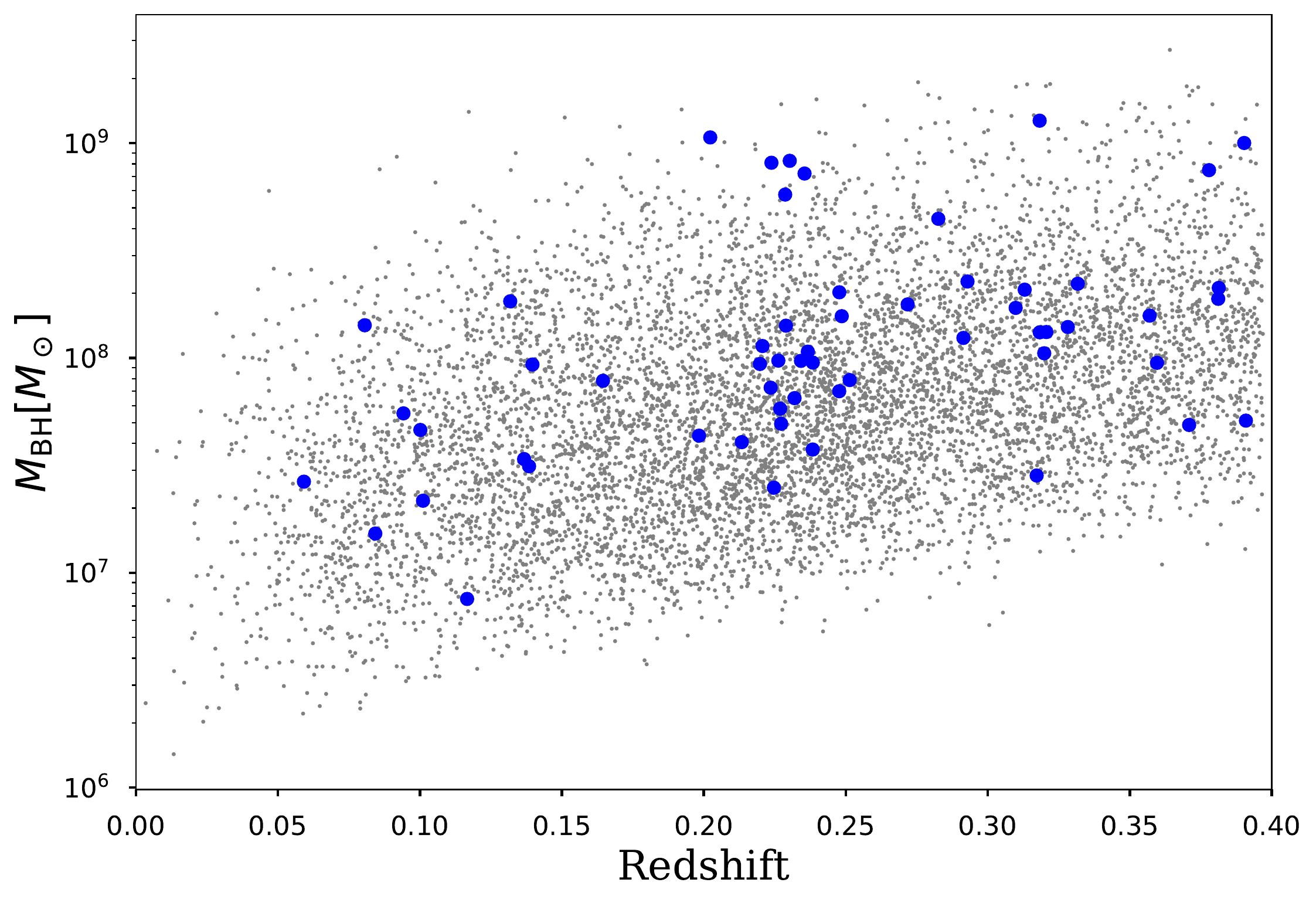}
    \caption{BH mass distribution as a function of redshift. The symbols are the same as in Figure~\ref{fig:Redshift}.}
    \label{fig:SMBHmass}
\end{figure}

Figure~\ref{fig:Redshift} shows the redshift distributions 
as a function of $\lbolothree$ and $\lbolir$ for the luminosity declining AGN candidates
(blue filled circle) and the parent sample (gray dot).
Figure~\ref{fig:Redshift} shows two important things.
One is that the median $\lbolir$ of luminosity declining AGN candidates are almost similar
to that of the parent sample, while the median $\lbolothree$ is one order of magnitude
higher than that of the parent sample. This indicates that our luminosity declining AGN candidates
had a very luminous AGN phase in the past and we will discuss this part later in Section \ref{chap:burst}.

The second is that our luminosity declining AGN candidates are slightly biased to higher redshift 
at $z>0.2$. This is partly due to the combination of our selection criteria 
for selecting large \othree~luminosities with $\lbolothree>10^{45}$~erg~s$^{-1}$
as shown in Figure~\ref{fig:boloplot},
and such sources are a dominant population only at $z>0.2$ as shown in the left panel of Figure~\ref{fig:Redshift}. Actually, 77\% of the luminosity declining AGN candidates are located in a redshift $>0.2$, which is a higher fraction compared to the parent sample (62\%).
This is also suggested from the difference in the redshift distribution of the two populations (luminosity declining AGN candidates and parent sample), showing the $p$-value of 0.03, suggesting that redshift distribution is slightly skewed to higher redshift for our luminosity declining AGN candidates, and the $p$-value becomes 0.08
 if we limit our sample only to sources with $z>0.2$, which suggests a non-significant difference in the distribution \citep{pvalue}.

\subsubsection{BH Mass and Bolometric Luminosity distributions}

The BH mass ($\mbh$), one key measurement obtained from the SDSS spectra, is also
compiled for all of our samples using the broad 
\halpha~emission lines \citep{HalpaSMBHmass}, through the spectral fitting done by 
\cite{MULLANEY} 
\footnote{
The equation of \cite{HalpaSMBHmass} allows
a maximum limit of reddening of $E(B-V)\sim0.12$ by following \cite{1989Cardelli}.
The extinction corrected \othree~luminosities 
sometimes reach unrealistically high \othree~luminosities for some AGN. This is a reason why we apply the observed broad \halpha~luminosity in this study. 
Note that using the extinction corrected broad \halpha~luminosity \cite{2013Dominguez,2000Calzetti}
for estimating the $\mbh$ would not change our main results because such a sample with unrealistically high AGN luminosity is not a dominant population.
}.

The both population of luminosity declining AGN candidates and the parent sample has a similar
median BH mass of $\left< \log (\mbh/\msun) \right> = 8.0\pm 0.5$ (for luminosity declining AGN candidates) and
$\left< \log (\mbh/\msun) \right>=7.8\pm 0.5$ (for the parent sample).
This suggests that our selection 
does not show a statistically significant difference on the BH mass as also seen in Figure \ref{fig:SMBHmass}.

\subsection{AGN Lightcurves in the last $\sim10^{4}$~yr}\label{sec:AGNlightcurve}

\begin{figure*}
    \centering
    \includegraphics[width=0.7\linewidth]{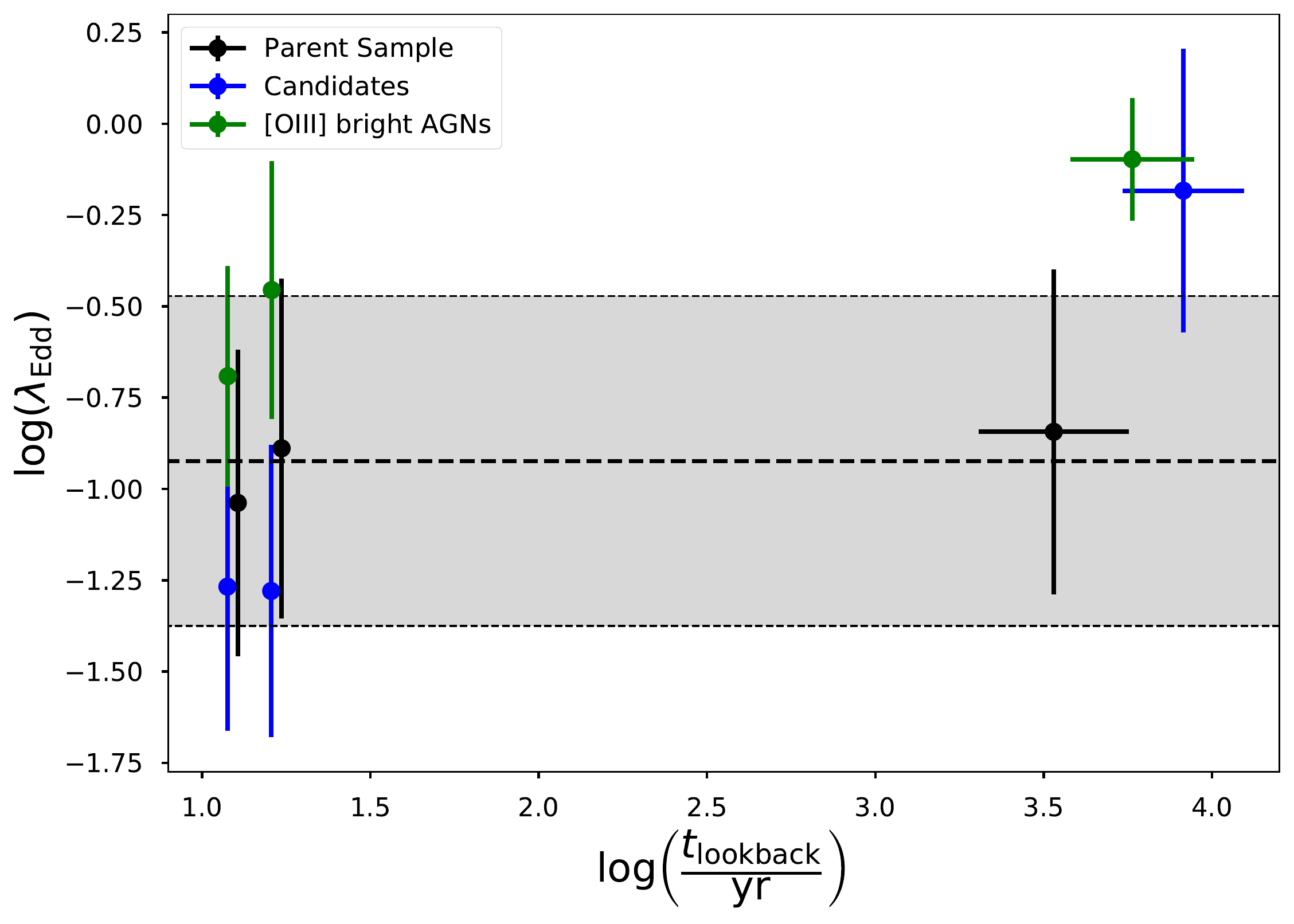}
    \caption{Eddington ratios as a function of the lookback time ($t_\mathrm{lookback}$) for luminosity declining AGN candidates, parent sample, and \othree~bright AGN. The three lookback times were calculated based on the averaged emitting size of the accretion disks, dusty tori, and the NLRs. 
    Here the dashed lines and grey area represent the median and scatter of the $\eddington$ of the parent sample.  The difference of the observation time between the SDSS DR7 and the \textit{WISE} observations are taken into account to estimate the $t_\mathrm{lookback}$, whose reference time is set on 2022 February 02 (MJD=59612).
  The $t_\mathrm{lookback}$ of the accretion disk and torus for the parent sample are shifted 0.03 dex to the right for clarity.
    }
    \label{fig:lookback}
\end{figure*}

One of our goals is to obtain how rapidly
luminosity declining AGN candidates have experienced
the luminosity decline in the last $\sim 3\times10^3$~yrs. 
To investigate this long-term AGN variability, we estimate the Eddington ratio ($\eddington \equiv L_\mathrm{bol} / \ledd$, where $\ledd$ is Eddington luminosity $\ledd \simeq 1.26 \times 10^{38} (\mbh/\msun)$~erg~s$^{-1}$) of our sample.
Considering that a few bolometric luminosities are estimated based on the
different AGN indicators tracing different physical scales as discussed in Section~\ref{chap:Sizecal}, we estimate the Eddington ratio for each AGN indicator.

Figure~\ref{fig:lookback} shows the time evolution of the
average Eddington ratio for the luminosity declining AGN candidates and parent sample in the last $\sim 3\times10^3$~yr.
The look back time ($t_\mathrm{lookback}$) was calculated
based on the light crossing time of the three AGN indicators with different physical scales
of accretion disk, dusty tori, and NLR (see Section~\ref{chap:Sizecal}).
We also take into account the different observation dates for the \textit{WISE} and SDSS observations
and we chose 2022 February 2nd (or MJD=59612) as our reference time ($t_\mathrm{lookback}$=0).
The average MJD of the \textit{WISE} observation is 55335/55325 for the luminosity declining AGN candidates/parent sample and the average one of the SDSS DR7 is 52997/53046. 
As a result of the different observation epochs between the \textit{WISE} and SDSS DR7, 
the average $t_\mathrm{lookback}$ for the accretion disk and dusty tori
is the almost same value as shown in Figure~\ref{fig:lookback}.

Figure~\ref{fig:lookback} indicates two important things for our luminosity declining AGN candidates and the parent sample.
One is that our parent sample shows on average the constant Eddington ratio over the entire time range, that is, the previous $\sim3\times10^3$~yrs,
indicating that the intrinsic variability should be on average smaller than the scatter of $\sim$0.4~dex.
This scatter value is consistent with previously known variability strength of typical AGN in the timescale of $\sim10$~yr \citep[e.g.,][]{1994Hook,2007Sesar,mcl10,2010Kozlowski},
and our results suggest that the stochastic variabilities
could be within the scatter of 0.4~dex even for the longer timescale of $\sim10^{3-4}$~yr. 
This long-term stability of AGN luminosity over the timescale of $\sim10^{3-4}$~yr is partially suggested from the relation between the X-ray and \othree~ luminosity correlations of AGN in the local universe, which shows a correlation with the scatter of $\sigma\sim0.4$~dex \citep{2006Panessa,2015Ueda,2015Berney}.

The second is that luminosity declining AGN candidates show a large luminosity decline between the \othree~emission region and the torus one, that is, in the previous $\sim10^{3-4}$~yrs. This is a natural outcome since we set the ratio cut of $R<0.1$. A more interesting trend of the light-curve for luminosity declining AGN candidates is that previous AGN luminosities traced by the \othree~emission region have reached almost Eddington limit with $\log \lambda_\mathrm{Edd} \sim 0$, while the current Eddington ratio is almost similar with those of the parent sample. 
This indicates that our fading AGN selection applied for type-1 AGN turn out to be an efficient way to find AGN who experienced the past AGN burst reaching the Eddington limit and this is partially suggested from Figure~\ref{fig:Redshift} and Section~\ref{sec:LAGN_vs_z}.

One might wonder that the bolometric correction for \othree~line might not be the same values between the standard disk and super-Eddington phase accretion. 
Although there are no studies on the bolometric corrections including super-Eddington phase,
 our luminosity declining AGN candidates (and also the parent type-1 AGN sample) do not 
contain extremely super-Eddington sources reaching $\eddington \gg 1$, with the maximum value of $\eddington\simeq 4$
 in the NLR.
 This means that even if the bolometric correction is not constant in the super-Eddington phase, the maximum difference is up to by a factor of 4,
 and the difference is likely smaller for most of the super-Eddington sources.
 Therefore, the constant bolometric correction used in this study does not strongly affect main results in this study.

One might also wonder that all Eddington-limit or super-Eddington sources in the \othree~emitting region might also have a similar trend of the AGN luminosity declining as seen in the luminosity decline AGN candidates.
To check this, we also selected \othree~bright AGN that are defined as $\eddington>0.5$
based on the \othree~based AGN bolometric luminosities.
This luminosity cut selects the sources with the average value of $\eddington=$0.8,
which is almost consistent with the one of luminosity decline AGN of $\eddington=0.7$.
The green crosses in Figure~\ref{fig:lookback} show the one for \othree~bright AGN
and they certainly show a slight luminosity decline in the last $10^{3-4}$~yr,
while the luminosity decline is smaller than that of luminosity decline AGN candidates.
This suggests that higher Eddington sources have a shorter lifetime residing in such a high Eddington ratio phase compared to the one in the lower Eddington ratio, but our luminosity decline AGN might have an additional mechanism to produce such a large luminosity decline reaching a factor of $>10$ in the timescale of $10^{3-4}$~yr.

\subsection{IR Variability tracing the past $\sim10$~yr}

\begin{figure*}
\begin{center}
\includegraphics[width=0.7\textwidth]{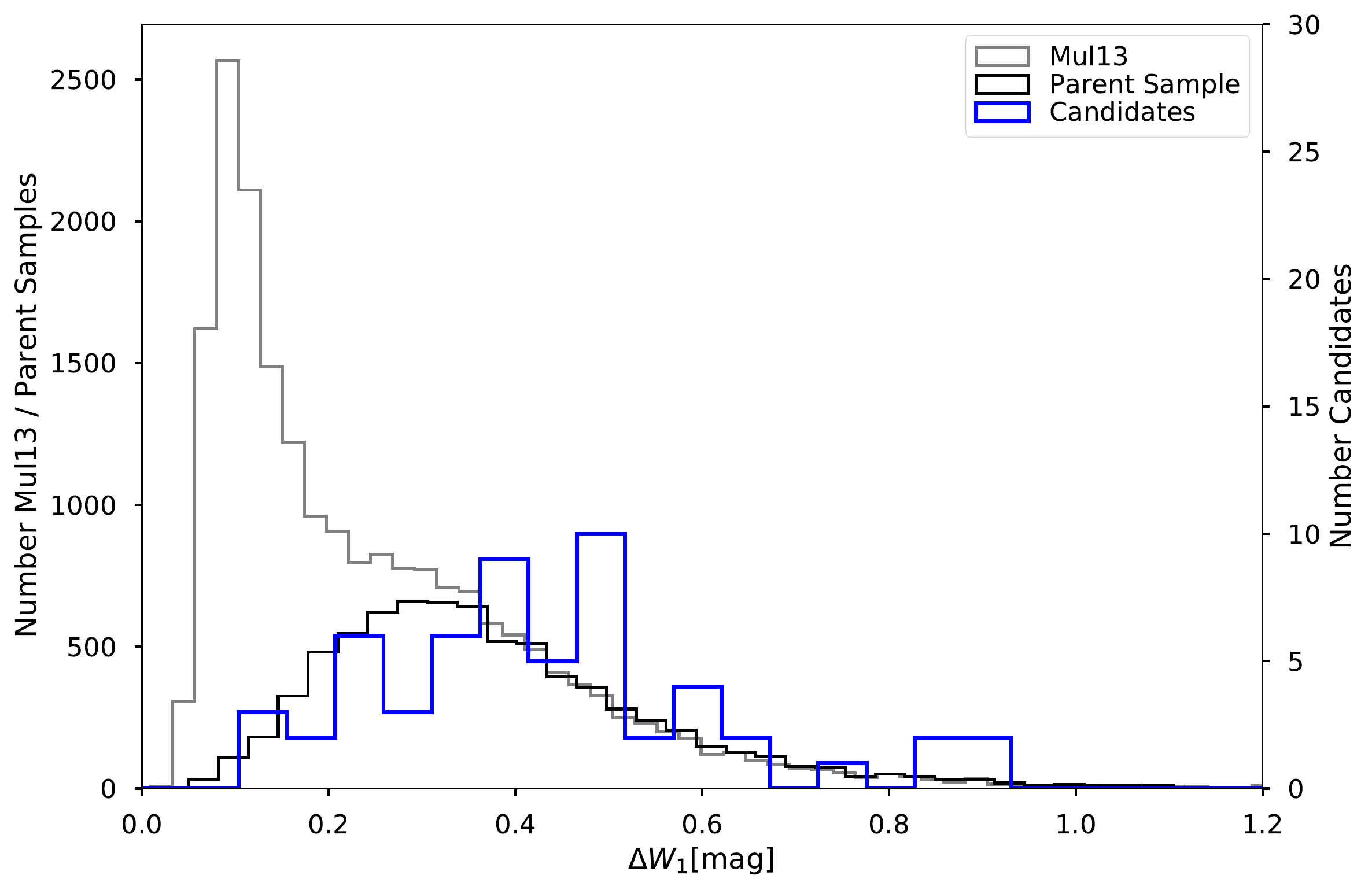}~
\caption{
Comparison of the variability distribution of the $\Delta W1$ for luminosity declining AGN candidates in blue, the parent sample in black, and additional to the complete AGN catalog of \citet[][Mul13]{MULLANEY}  containing also type-2 AGN in grey. 
The left y-axis shows the number of the sources for the parent sample and the catalog of \citet[][Mul13]{MULLANEY} and the right axis shows for the luminosity declining AGN candidates.
}\label{fig:IRVar}
\end{center}
\end{figure*}

It is worthwhile to trace the IR luminosity variability from the dusty torus
by using the \textit{ALLWISE} and \textit{NEOWISE}  IR light curves of W1 and \textit{W2}-bands covering the last 10~yr \citep{NEOWISE_Mainzer14,NEOWISE_DOI} and they are suitable bands for tracing the relatively inner part of the AGN dusty torus emission \citep[e.g.,][]{ste12,2012Mateos,2018Assef}.
The \textit{ALLWISE} database provides  the \textit{WISE} All-Sky catalog, the \textit{WISE} 3-Band Cryo  and \textit{WISE} Post Cryo catalog which are covering the observation from December 14, 2009 -
September 29, 2010 \citep{WISE_WRIGHT10,WISECRYO_DOI,PostCryoWISE_Mainzer11}.  The \textit{NEOWISE} data covers additional IR multi-epoch data for \textit{W1} and \textit{W2} bands
and we utilized \textit{NEOWISE}  2021 datarelease which contains multiepoch photometries between December 13, 2013 and December 13, 2020 \citep{PostCryoWISE_Mainzer11,NEOWISE_Mainzer14}.
\textit{WISE} has a 90 minute orbit and conducts $\sim14.1$~average observations for the parent sample and $\sim 14.7$~average observations for the luminosity declining AGN candidates
over a $\approx1$~day period, and a given
location is observed every 6 months,
which means that one source has on average $16 \times 14$--$15$ data points for a light curve spanning $\sim10$~yr long.

In this study, we applied the cross-matching radius of 2~arcsec with
the SDSS coordinates of our sample. 
Out of the two bands, we used only the \textit{W1}-band because 1) \textit{W1}-band traces
a warmer dusty region of $T\sim 900$~K than that traced by \textit{W2}-band, that is, 
a smaller physical scale and a corresponding response
of the accretion disk variability is shorter
and 2) the \textit{W1}-band is more sensitive than \textit{W2}-band, which enables us to extract more sources
even in a single exposure.

We obtained the \texttt{w1mpro} photometry,
and selected the photometries with the good flux quality
with \texttt{ph\_qual=A} or \texttt{=B}, which selected the sources with SN$>3$,
and also the photometries with no flux contamination by using \texttt{ccflag=0}.
Based on the selection criteria above,
we obtained the \textit{WISE} W1 IR light curves for 7,614 (parent sample) and all 57 sources
(luminosity declining AGN candidates) out of the 7653 and 57 sources, respectively.

We binned the cadence data shorter than one day
and derived the median magnitude,
and we call each longer cadence observation a single epoch of observations.
This means that there are typically 16 epochs of photometry
with separations from 6~month to a maximum of nearly $\sim10$~yr.
The average detections per epoch is $\sim 14$ for the parent sample.
Sometimes there are fewer detections in each epoch,
and we used the epoch which has at least two detections with good photometric quality flags.
We then followed the same manner of \cite{Stern18},
who obtained the maximum and minimum magnitudes 
($W1_\mathrm{max,min}$ where max,min stands for the maximum and minimum magnitudes) from the multiple
epoch observation of each source, and calculated 
the variability strength
$\Delta W1=|W1_\mathrm{max}-W1_\mathrm{min}|$.

Figure~\ref{fig:IRVar} shows a histogram of the sources as a function of $\Delta$W1.
The median variation of parent sample is $\left< \Delta\mathrm{W1} \right>=0.35\pm 0.18$, which is significantly higher than that of \cite{Stern18} with $\left< \Delta\mathrm{W1} \right>\sim 0.2$, because of the difference of the sample and the resulting inclination angle effect of the dusty tori.
While our sample is purely type-1 AGN with the face-on view of the central engine, 
the sample of \cite{Stern18} also contains type-2 AGN based on their nature of the \textit{WISE} IR selection, and those type-2 AGN would show less variation in $\Delta$W1 since the emitting region of the \textit{W1}-band is close to the sublimation region of the dusty region \citep[e.g.,][]{kos14}, and dust obscuration reduces the observed variability.
We also checked this tendency by including the type-2 AGN sample of \cite{MULLANEY}, and we obtained the almost similar median values of $\Delta$W1, which is shown as a gray histogram in Figure~\ref{fig:IRVar}.

Figure~\ref{fig:IRVar} also shows a different distribution of $\Delta$W1 between the parent sample and luminosity declining AGN candidates, which shows a relatively larger $\Delta$W1, 
whose p-value for these two populations is 0.02, implying a statistically significant difference in the samples. 
The median value for the luminosity declining AGN candidates is $\Delta\mathrm{W1}=0.41\pm0.18$, which is larger than that of the parent sample ($\Delta\mathrm{W1}=0.35\pm0.18$).
This is suggestive that our luminosity declining AGN candidates are on average more IR variable
rather than the parent sample not only in the timescale of $\sim10^3$~yr but also even in the timescale of $\sim10$~yr, which is a corresponding timescale of changing-look AGN \citep[e.g.,][]{Stern18}.

\begin{figure*}
\begin{center}
    \includegraphics[width=0.5\linewidth]{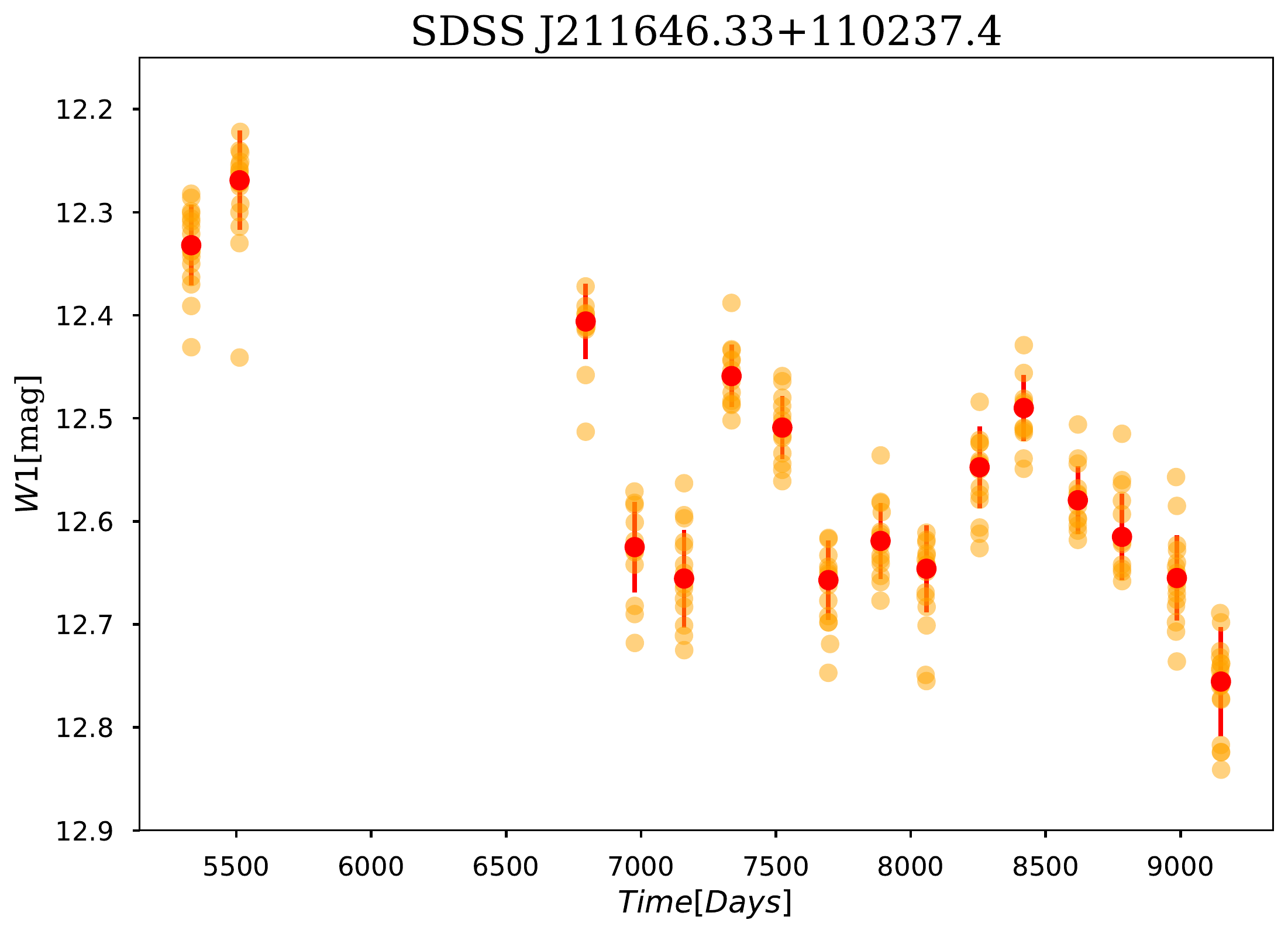}~
    \includegraphics[width=0.5\linewidth]{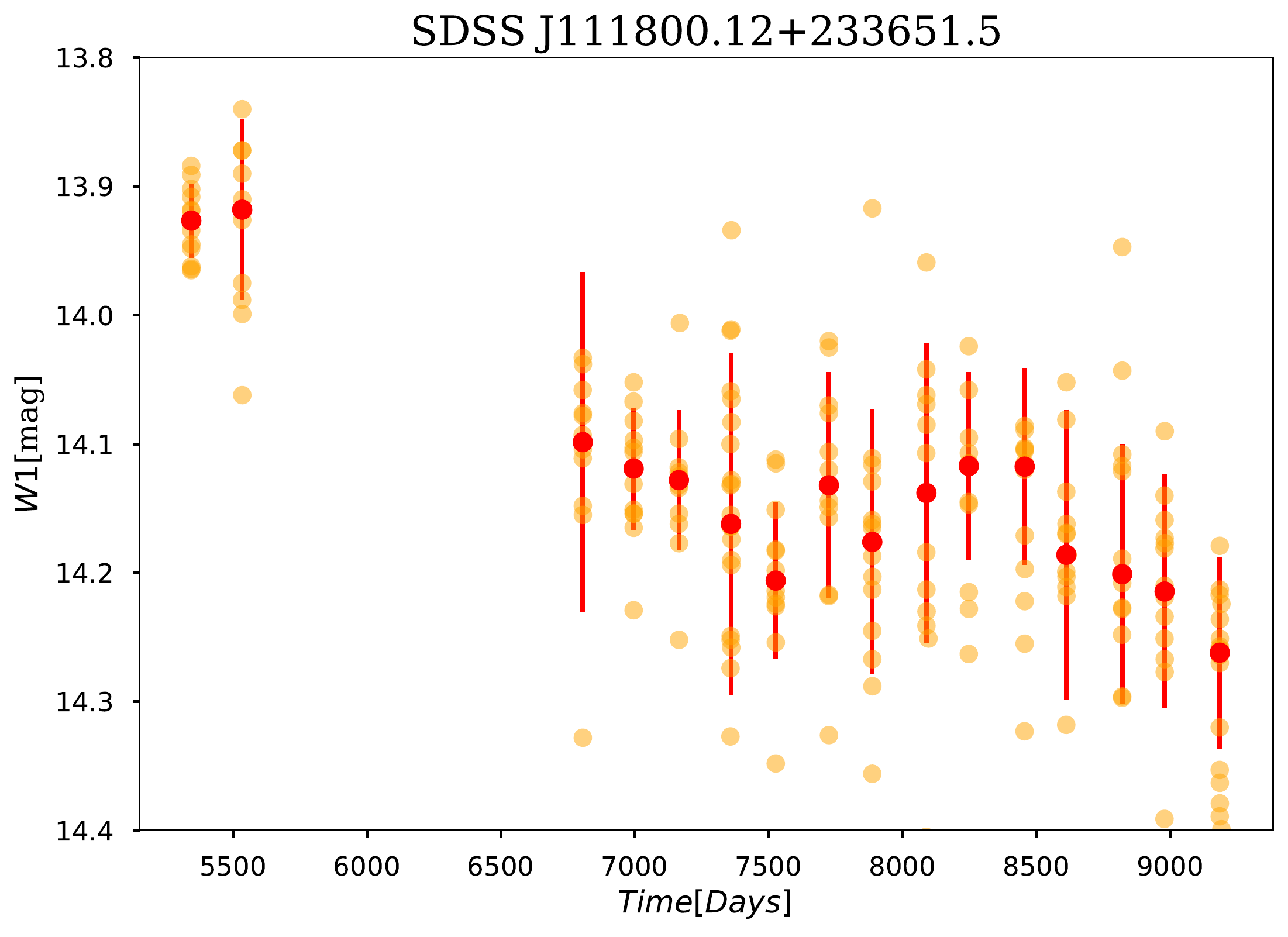}~
   \caption{\textit{WISE} and \textit{NEOWISE} light curves at 3.4~$\mu$m of two of the luminosity declining AGN candidates, SDSS J211636.33+110237.4 (left) and SDSS J111800.12+233651.5 (right). Both light curves show a continuous flux decrease over the 10~yr long time span with $\Delta W1\sim 0.49$ mag (left) and $\Delta W1\sim 0.34$ mag (right), respectively.}
    \label{fig:NeoFading}
\end{center}
\end{figure*}

The \textit{W1}-band lightcurves of the 57 luminosity declining AGN candidates show a variety of their IR variabilities; continuous increasing, decreasing, and a stochastic trend.
%
Four sources show a significant flux increase in the \textit{W1}-band, which might suggest a recovery from the slightly lower accretion. 
On the other hand, two sources show a 
clear continuous flux decline over $\sim10$~yr. These two lightcurves are shown in Figure~\ref{fig:NeoFading} (left, SDSS J211646) with $>\Delta W1\sim 0.49$~mag and (right, SDSS J111800) with $>\Delta W1\sim 0.34$~mag, which suggests a continuous fading of the AGN dust emission over the past 10 yrs. 
The change of the Eddington ratio for those two sources are from $\eddington = 0.03$ to $\eddington=0.02$ for SDSS J211646 and 
$\eddington = 0.16$ to $\eddington=0.12$ for SDSS J111800.
The remaining 51 sources show stochastic variabilities that are not classified as either of the continuous declining or increasing.


\subsection{Multi-Epoch SDSS Optical Spectra Spanning $1$--$10$~yr Time Gap}\label{sec:fitting}
\begin{figure*}
\begin{center}
    \includegraphics[width=0.5\linewidth]{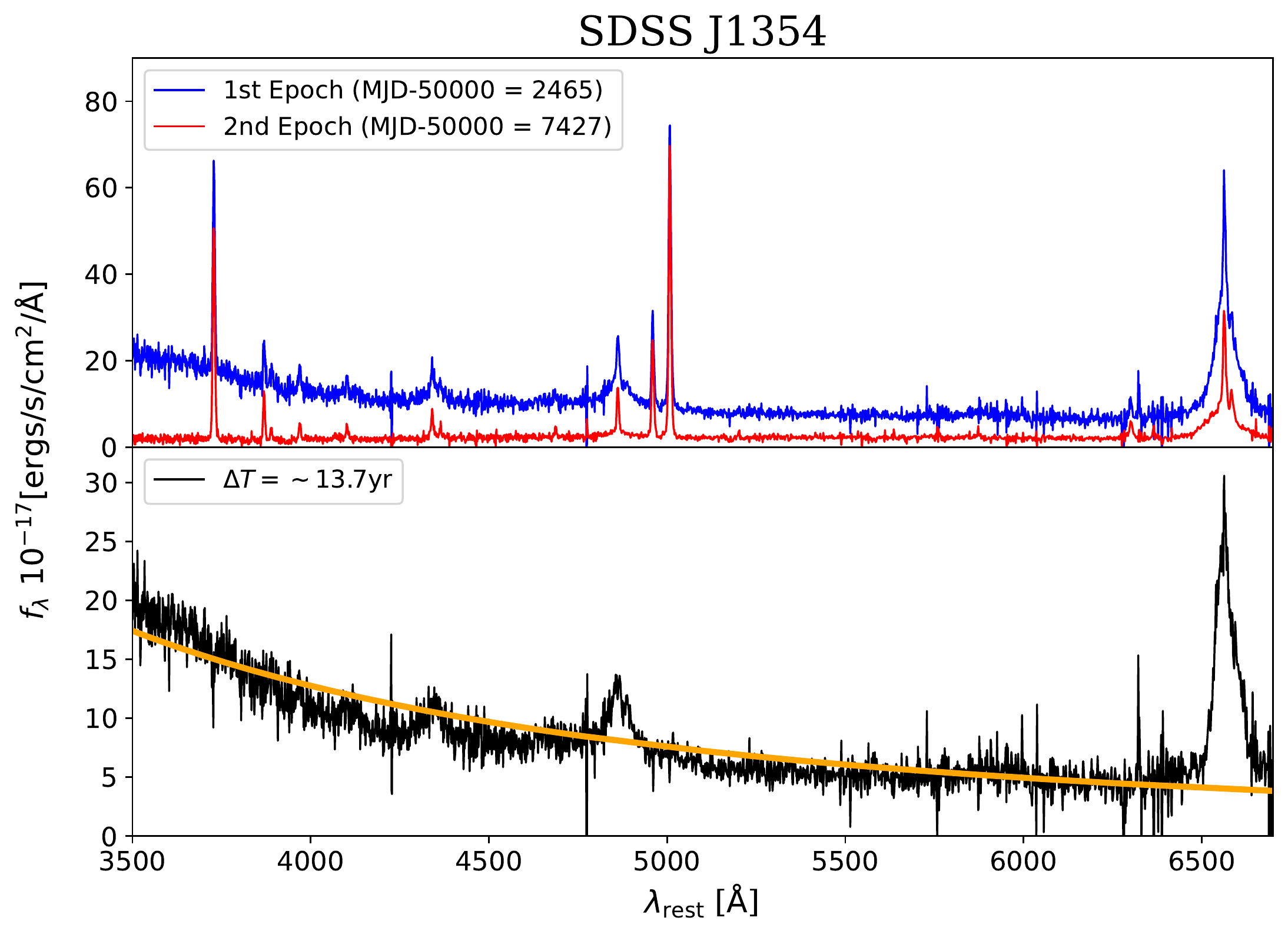}~
    \includegraphics[width=0.5\linewidth]{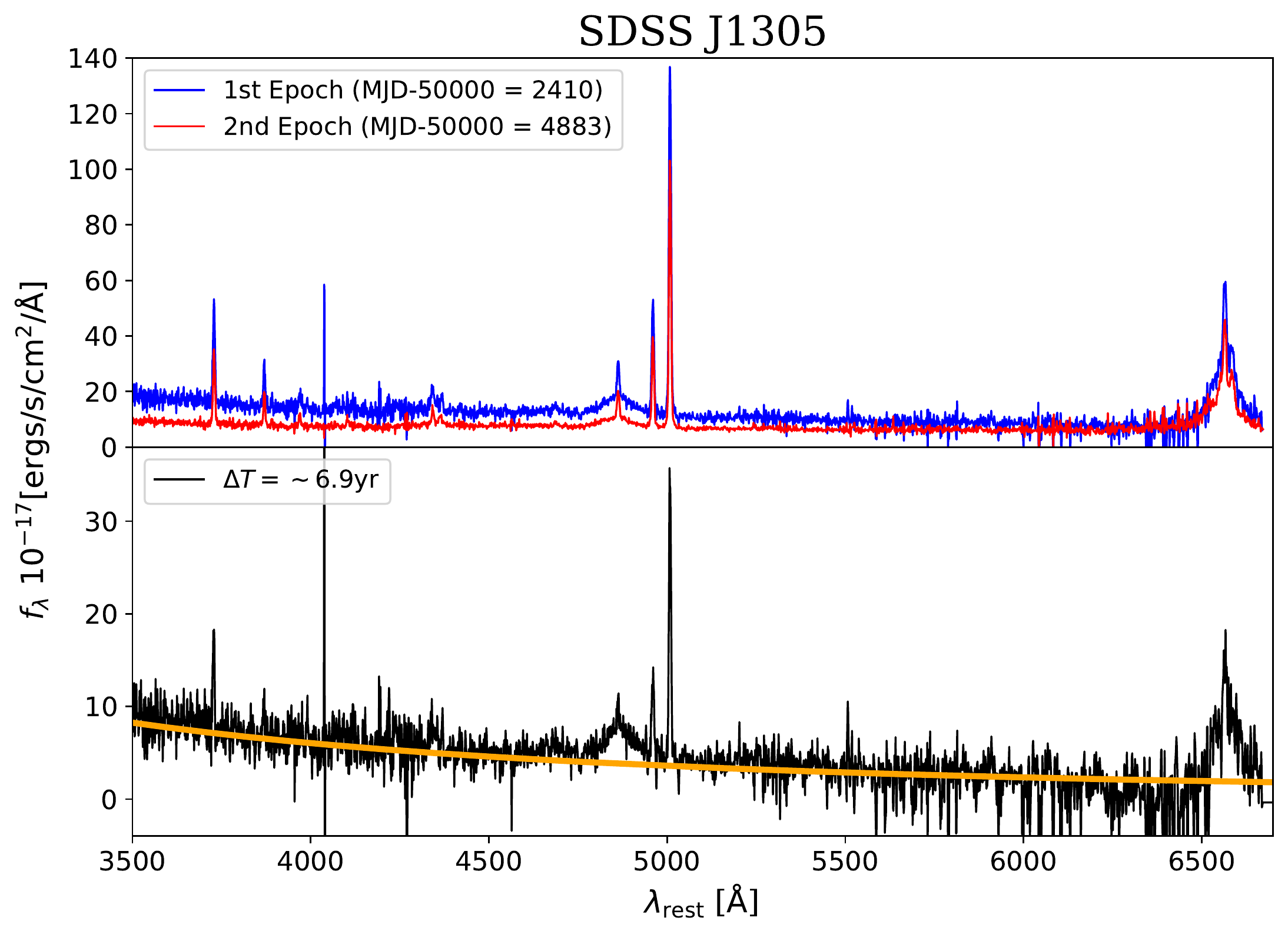}~\\
    \includegraphics[width=0.5\linewidth]{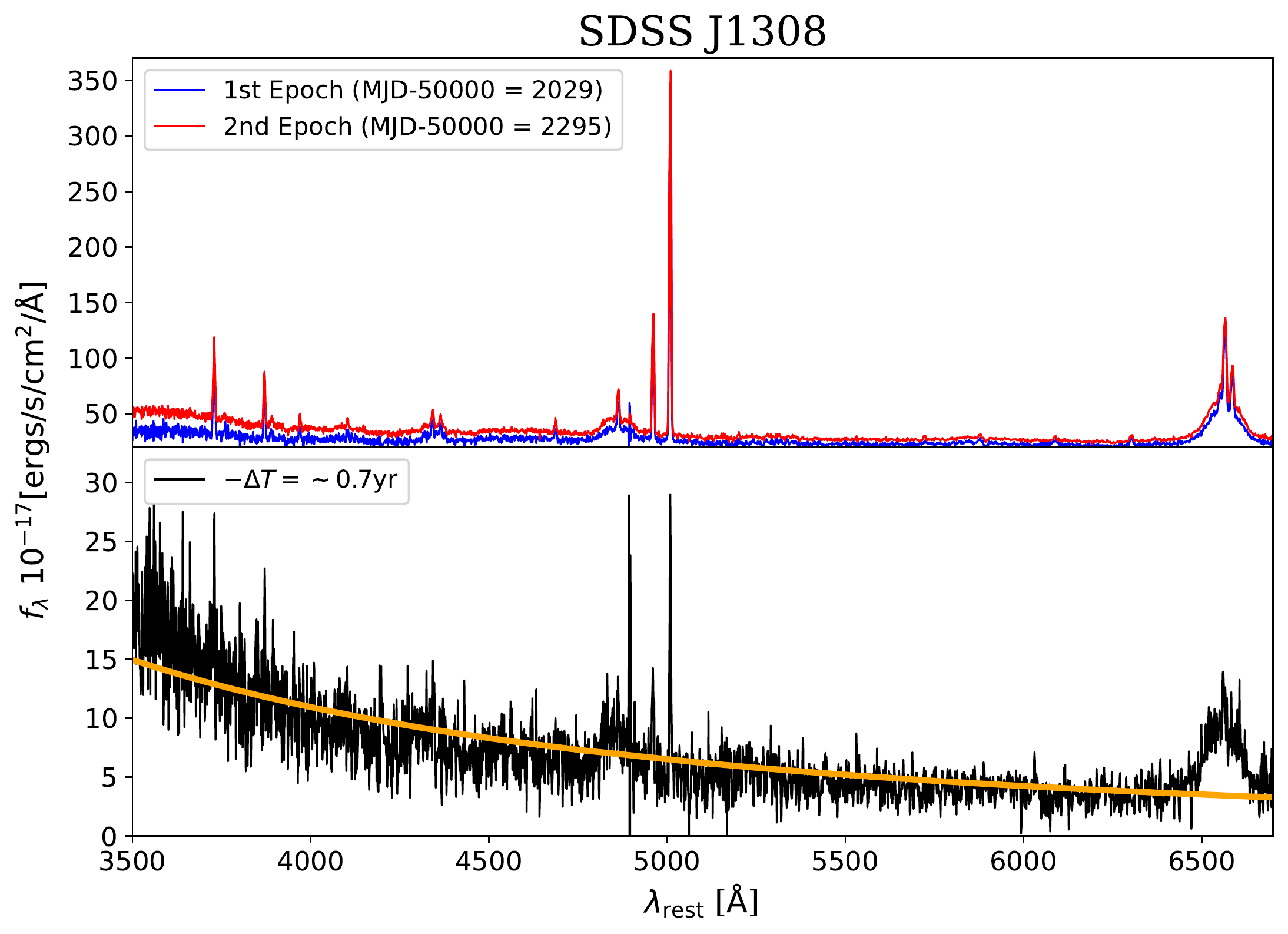}~
    \includegraphics[width=0.5\linewidth]{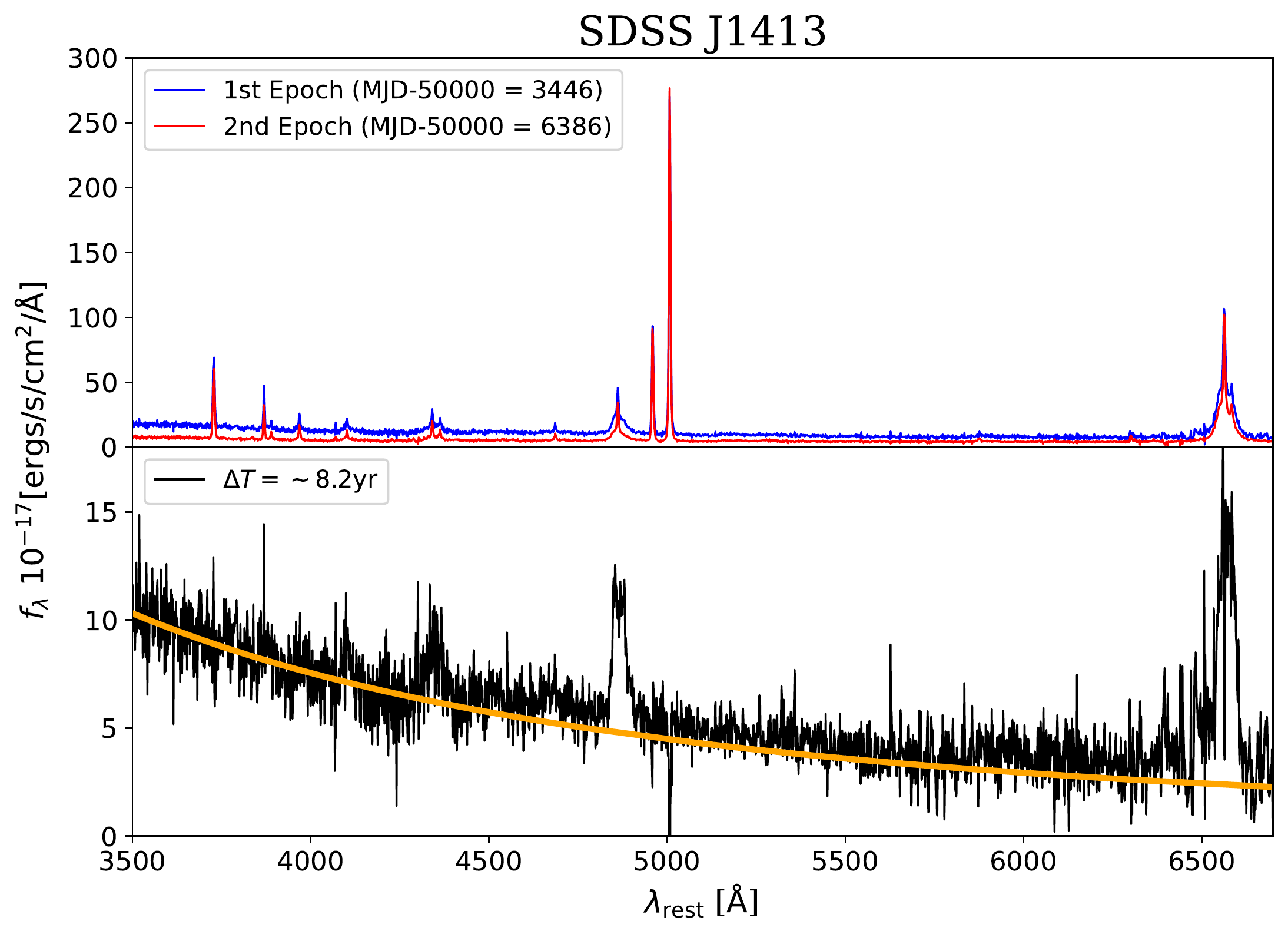}~\\
    \caption{Top panels: The rest-frame optical spectra of the luminosity declining AGN candidates showing a changing-look behavior between the two epochs of the SDSS observations (the blue and red line represents the first and second epoch one, respectively). The observation dates are shown with the unit of MJD in the legend of each panel.
    Bottom panels: The subtracted spectrum between the two epoch (black) overlaid with the fitting curve of the theoretical accretion disk of $f_\lambda\propto \lambda^{2.3}$ (orange)  \citep{1973Shakura}.}
    \label{fig:Obj_Spec}
\end{center}
\end{figure*}

The $\sim10$~yr long IR variability by the \textit{WISE} \textit{W1}-band indicates that the luminosity declining AGN candidates tend to show
large variabilities both in $\sim10$~yr and $\sim10^3$~yr.
This motivates us to search the multi-epoch spectra given that
our sample is obtained from the SDSS spectral surveys.
We searched for multi-epoch spectra for our luminosity declining AGN candidates.
Out of the 57 luminosity declining AGN candidates, 13 sources have the second epoch spectra observed in the later SDSS eBOSS survey \citep{SDSSDR7,SDSSBOSS,SDSSSEGUE,SDSSeBOSS}.
The time difference between the first and second epoch spectra spans from $\sim$1 to 14~yr.

Considering that the fiber sizes are different 
between the SDSS BOSS (2~arcsec) and SDSS eBOSS (3~arcsec),
the host galaxy continuum and the emission might contaminate strongly for the later SDSS eBOSS spectrum.
On the other hand, this does not effect our study significantly, since our main motivation is to compare the central accretion disk components and the broad emission lines, both of which are considerably smaller than the current SDSS aperture sizes and
their emission contributes equally to both of the spectra.

Based on the subtraction of the two-epoch spectra, 
we found four sources with a significant difference 
of the continuum and/or broad emission lines.
Top panels of Figure \ref{fig:Obj_Spec} shows two AGN with a clear continuum and the broad component difference between the two epochs,
so called changing-look (CL) AGN
\citep{2015LaMassa,MacLoadCLSearchPhoto,Stern18}.
SDSS J1354 shows a disappearance of the
blue excess continuum associating the accretion disk emission and H$\beta$ broad component,
and a weakened H$\alpha$ broad line,
resulting a spectral type change from type-1 into type-1.9 \citep{1976Osterbrock,1977Osterborck,1981Osterbrock}.
SDSS J1305 shows a disappearance of the
blue excess continuum and a weakened H$\beta$,
a spectral type change from type-1 into type-1.8.
We also summarized the spectral properties in Table~\ref{tab:CLinfo}.

The bottom panels of Figure \ref{fig:Obj_Spec} shows two AGN with a significant continuum flux change (appearance for SDSS J1308 and disappearance for SDSS J1413) stronger than $3\sigma$ error of the associating continuum, but do not show a difference in broad emission components. We hereafter call those two sources as changing-look behavior (CLB) AGN
to distinguish from the CL AGN with a clear sign of the flux changes of the broad emission.

The detection rate of CL AGN for our luminosity declining AGN candidates is extremely high reaching 15\% ($2/13$), and even higher of 30\% ($4/13$) if CLB AGN are included. This is a much more efficient pre-selection method by one to two orders of magnitude than the previous CL AGN searches, 
which have a range of detection rates from
0.4 to 1.3\% after the pre-selections.
For example, \cite{POTTSCL2021} selected the candidates by applying the $\Delta g_\mathrm{AB}>0.16$,
and found six sources out of the pre-selected 941 sources, whose success rate is $\sim0.6$\%.
\cite{yan18} also utilized the SDSS multi-epoch spectra
where one epoch is shown as a source category of ``GALAXY'', but the other epoch shows a different category of ``QSO'' (or vice-versa). This method pre-selected 2023 candidates and 9 sources were genuinely CL sources, whose success rate is $\sim0.4$\%.
The highest success rate was obtained through the study by \cite{MacLoadCLSearchPhoto},
who selected the candidates by applying the multi-epoch optical magnitude difference of
$|\Delta g_\mathrm{AB}|>1$~mag with good signal-to-noise of $\sigma_g < 0.15$~mag.
This method selected 1011 candidates and 10 sources were CL AGN, and resulting success rate is 1.0\%.
This indicates that the long-term variability selection used in our study turn out to be an efficient method to search CL AGN tracing the variabilities in $\sim10$~yr timescale.

\begin{figure*}
    \centering
    \includegraphics[width=8cm]{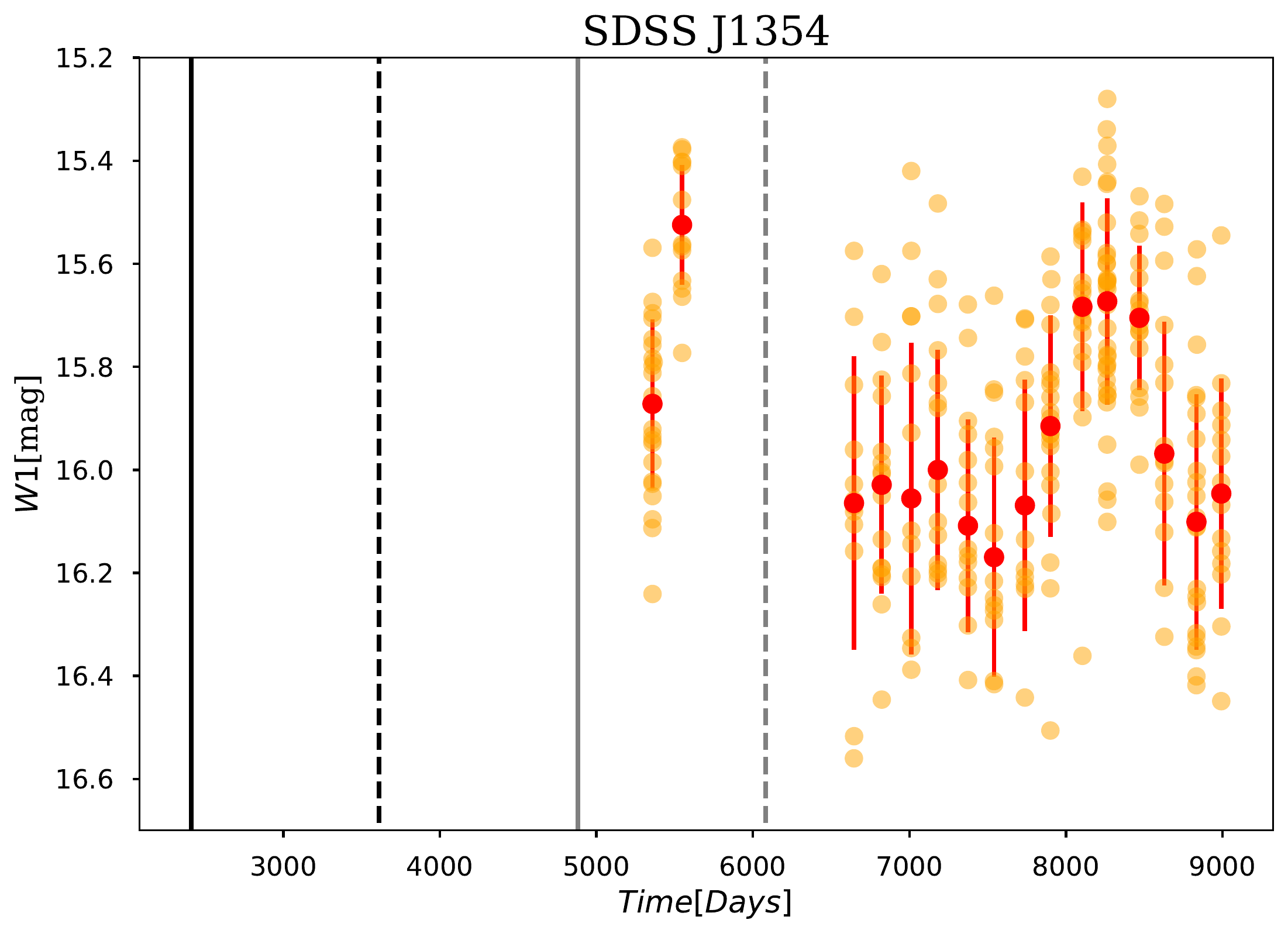}
    \qquad
    \includegraphics[width=8cm]{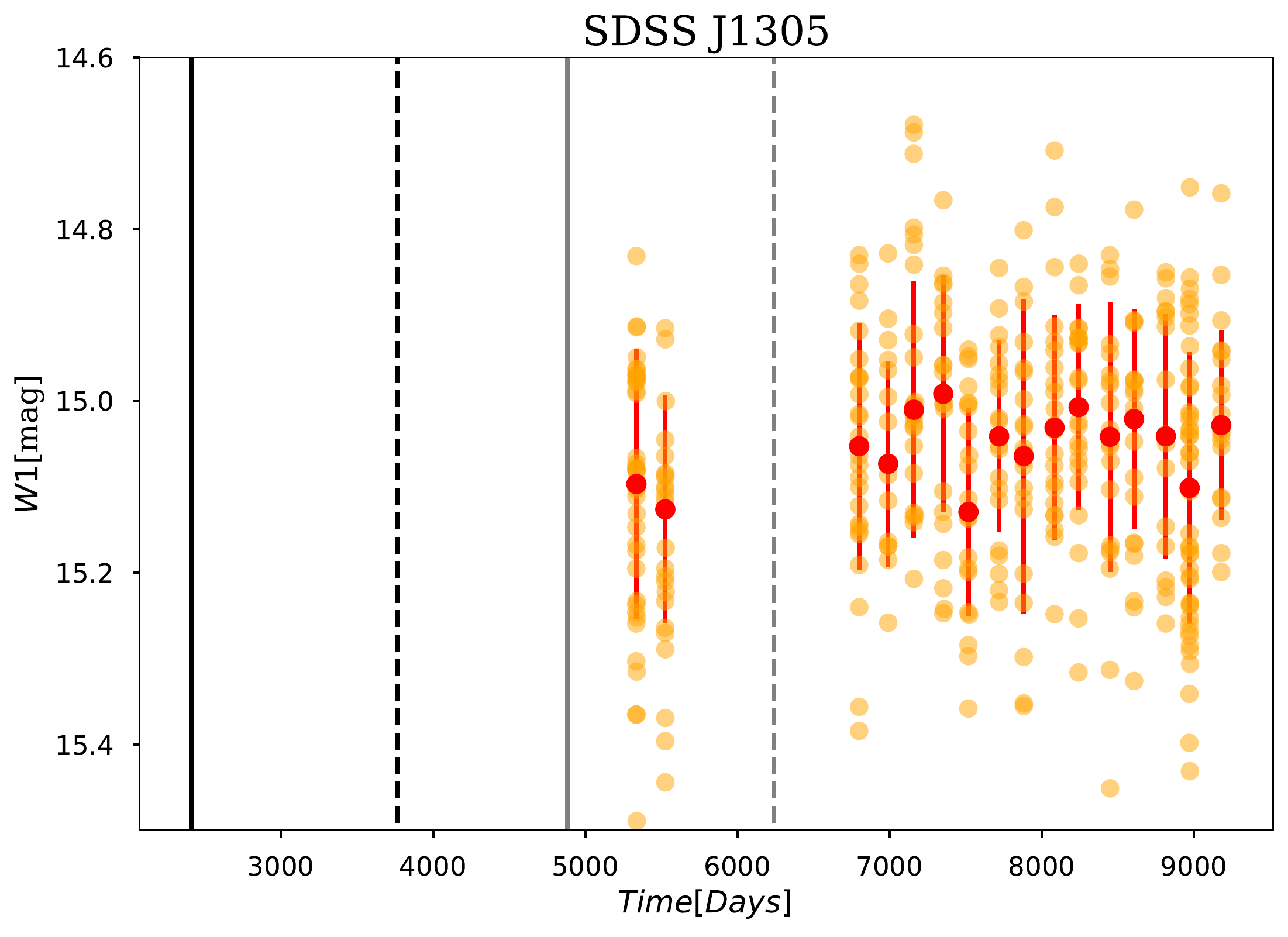}
    \\
    \includegraphics[width=8cm]{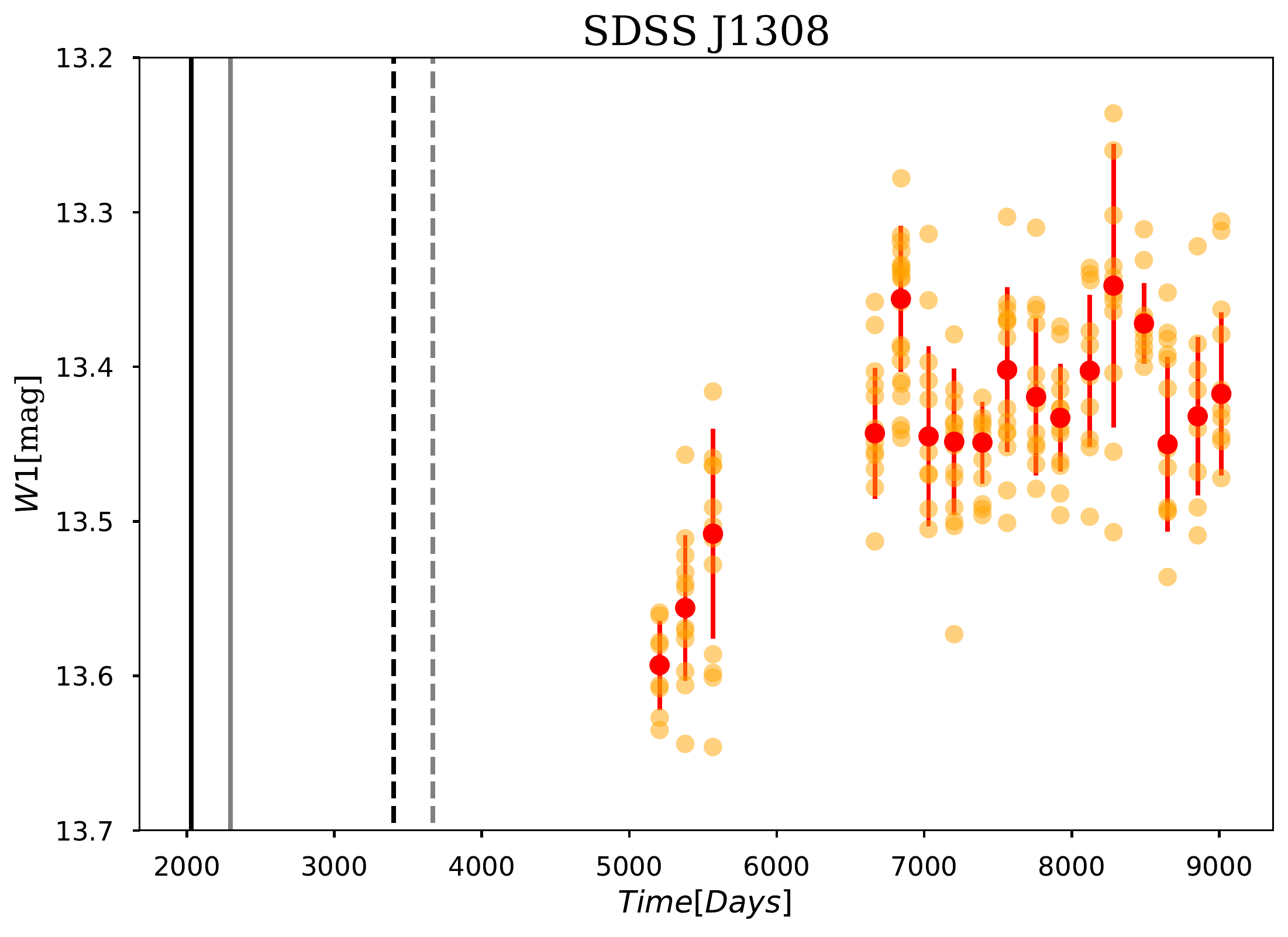}
    \qquad
    \includegraphics[width=8cm]{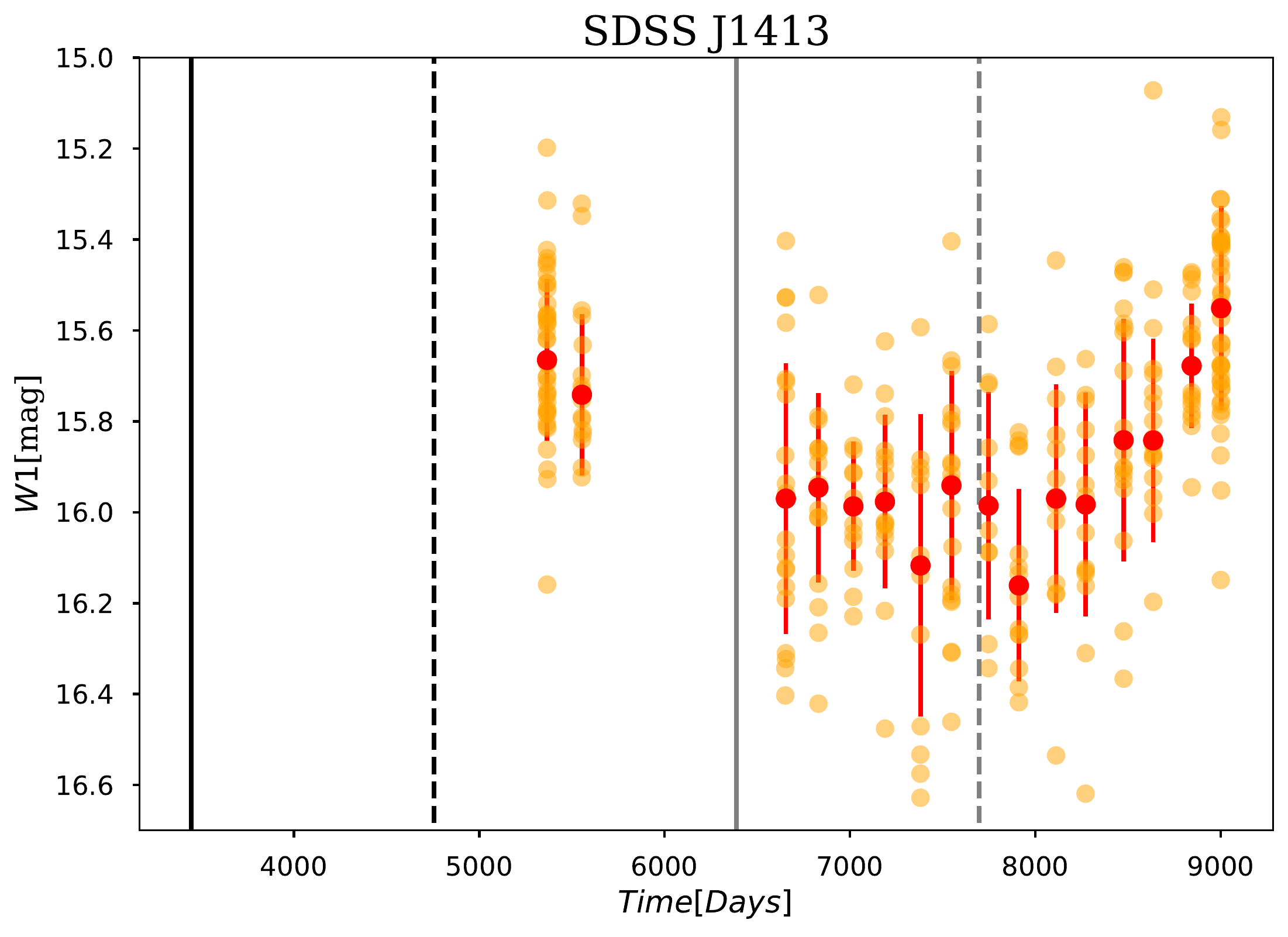}
    \caption{3.4~$\mu$m WISE+\textit{NEOWISE}  light curve of the four CL and CLB AGN. The orange filled circle represents the obtained W1 magnitude at a given observing time. The red filled circle with errorbar represents the binned W1 magnitude at each epoch. The vertical straight lines represent the observing date of the obtained SDSS spectra. The black/gray one represents the date of the first/second epoch spectrum, respectively. The dashed lines represent the time of the arrival of the information of the accretion disk into the dusty torus region traced by the \textit{W1}-band.}
    \label{fig:NeoCurve}
\end{figure*}

Figure~\ref{fig:NeoCurve} shows the \textit{ALLWISE} and \textit{NEOWISE} light curves
of the newly discovered four CL and CLB AGN.
The black and gray solid vertical lines in each panel represent
the observed MJD of the SDSS spectra, respectively.
The black and gray dashed vertical lines represent the 
expected arrival time of the AGN activity seen in the SDSS spectra at the emitting region of the \textit{WISE} \textit{W1}-band emission by considering the time delays between the accretion disk and \textit{WISE} \textit{W1}-band.
To estimate the light travel time from the center to the emitting region of the \textit{W1}-band, we utilize the relation of \cite{1987Barvainis} 
with the additional $k$-correction as following;
\begin{align}
    D_\mathrm{sub}/\mathrm{pc}=1.3\left(\dfrac{L_\mathrm{bol,AD}}{10^{46}\mathrm{~erg~s}^{-1}}\right)^{0.5}\left(\dfrac{T(1+z)}{1,500\mathrm{K}}\right)^{-2.8}
\end{align}
where we assume the emitted blackbody emission produced by the dust temperature of $T=850$K were detected in the \textit{WISE} \textit{W1}-band. 

Those observing epochs of SDSS spectra as shown in Figure~\ref{fig:NeoCurve} give a sign that, for SDSS J1413, the second epoch of the SDSS spectra
was taken at the faintest phase of the light-curve in the \textit{NEOWISE},
which is consistent with the ``disappearing'' sign in the SDSS spectra.
In addition, the current \textit{NEOWISE} light-curve already reaches the same flux level as the \textit{ALLWISE} fluxes which are close to the first epoch of the SDSS spectra, suggesting that the current optical spectra might show appearing feature again.
For SDSS~J1308 which shows an appearing feature, even though the two spectra are taken well before the \textit{ALLWISE} light-curve range, the burst or bright-end phase might last until the current epoch.

\begin{deluxetable*}{cccccccc}
\tabletypesize{\footnotesize}
\tablecolumns{7}
\tablewidth{0pt}
\tablecaption{Basic information about the four new found CL and CLB AGN.\label{tab:CLinfo}}
  \tablehead{
      \colhead{Name}&\colhead{SDSS OBJID DR7}&\colhead{Ra [°]} & \colhead{Dec [°]} & \colhead{$z$} &\colhead{Changing-look feature} &\colhead{Type} }
     \startdata
SDSS J135415.53+515925.7 & 587733410444607498 & 208.56 &  51.99 & 0.32 & turning off & CL\\
SDSS J130501.41+604204.9 & 587732592255434858 & 196.26 & 60.70 & 	0.38 & turning off & CL\\
SDSS J130842.24+021924.4 & 587726015076433951	& 197.18 &	2.32 & 0.14 & turning on  & CLB\\
SDSS J141324.21+493424.9 &588017713123557439	& 213.35	& 49.57	& 0.37 & turning off& CLB
\enddata
\end{deluxetable*}

\cite{Sheng17} showed that CL AGN show higher variabilities in the \textit{NEOWISE}  \textit{W1}-band with $\Delta \mathrm{W1}>0.4$~mag, which is consistent with the cases of SDSS J1354  and the CLB AGN SDSS J1413 with $\Delta \mathrm{W1}=0.65$ and $\Delta \mathrm{W1}=0.61$ mag, respectively.
On the other hand, the \textit{NEOWISE}  and the SDSS two epoch spectra do not simultaneously cover the transient phase of CL AGN for the remaining CL AGN and CLB AGN, and actually, they show smaller W1 variabilities of $\Delta \mathrm{W1}=0.14$ for J1305 and $\Delta \mathrm{W1}=0.25$ for J1308, respectively. This result indicates that the combination of the SDSS two epoch spectra and \textit{WISE} variability further increases the probability to find such CL AGN by expanding the variability timespan by a factor of $\sim2$ ($\sim 7000$~days).

\subsection{Location of luminosity declining AGN candidates in the BPT Diagram}

\begin{figure}
    \centering
    \includegraphics[width=\linewidth]{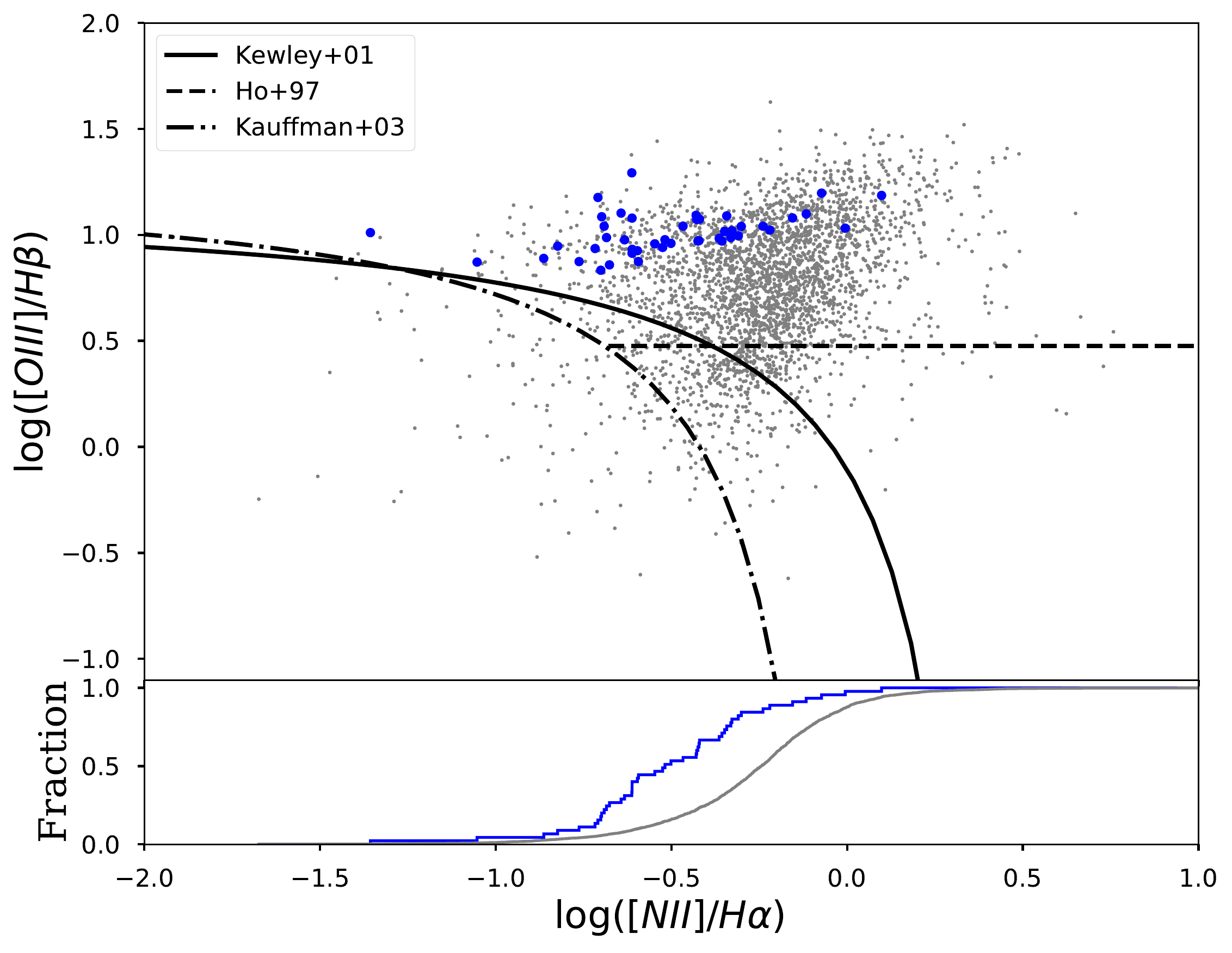}
    \caption{
    (Top): The distribution of luminosity declining AGN candidates (blue points) and parent sample (gray points) in the BPT diagram \citep{BPT_Baldwin81}. The two sequences are shown separating the AGN and starforming galaxies \citep{Kewley01,Kaufmann03} and the horizontal line separating the AGN and the LINERs \citep{Ho97Linercut}.
    Bottom panel: Cumulative histogram
    of the line flux ratio of log(\ntwo/\halpha).}
    \label{fig:BPT}
\end{figure}

Figure~\ref{fig:BPT} shows the location of 
our luminosity declining AGN candidates (blue circles)
and the parent type-1 AGN sources (gray dots) in the BPT diagram plane.
Here, we limit our AGN to the additional SN cut for the emission lines, with SN$>=5$ for the narrow \halpha, \hbeta~, and \ntwo~components,
which are necessary for the sources to plot in the BPT diagram. 
This reduces the sample into 44 luminosity declining AGN candidates and 3042 parent sources.

Figure~\ref{fig:BPT} demonstrates two important features of luminosity declining AGN candidates.
One is that luminosity declining AGN candidates preferentially have higher $\log (\mathrm{[OIII]/H\beta})$ ratio with $\log (\mathrm{[OIII]/H\beta}) \gtrsim 1.0$, which is likely originated from our selection criteria, requiring high observed \othree~ luminosity with $\log \lbolothree>10^{45}$~erg~s$^{-1}$ as shown in Figure~\ref{fig:Redshift}.

The second is that luminosity declining AGN candidates are located wider range of log(\ntwo/\halpha), spanning with 
$-1.5<$log(\ntwo/\halpha)$<0$, but is located preferentially in the low ratio of log(\ntwo/\halpha) $<-0.5$.
The median ratio of log(\ntwo/\halpha) is $-0.52\pm 0.27$ for the luminosity declining AGN candidates and $-0.24\pm 0.27$ for the parent sample.
The cumulative histogram in the bottom panel of Figure~\ref{fig:BPT} shows a notable difference between the two populations, with the 
$p$-value of $\sim 2 \times 10^{-8}$, showing a significant distribution difference.

\cite{2006Groves} showed that the line flux ratio of log(\ntwo/\halpha) depends strongly on the NLR gas
metallicity and \cite{2017Kawasaki} also demonstrated that
such low log(\ntwo/\halpha) ratio with log(\ntwo/\halpha)$=-1$--$-0.5$
cannot be described by the high ionization parameter alone, which is estimated from the oxygen line ratio diagrams of \othree/\otwo~and \oone/\othree.
They defined a low-metallicity AGN if the sources have a flux ratio of log(\ntwo/\halpha)$=-1$--$-0.5$, because the nitrogen relative abundance is in proportion to the metallicity, and roughly half of the luminosity declining AGN candidates fulfill such criterion.
This suggests that some of our luminosity declining AGN candidates might be in an early chemical metal enrichment phase of galaxies \citep{2017Kawasaki}. 

 \cite{2017Kawasaki} defined a ``BPT-valley'' AGN, which are selected based on the region above the sequence of \cite{Kaufmann03} (black solid curve) and the low Nitrogen abundance of log(\ntwo / \halpha) $ <- 0.5$. They found that a significant fraction of these selected AGN show low metallicity features.
We followed the same manner of \cite{2017Kawasaki} to select these BPT valley sources from our parent type-1 AGN sample, which leaves 721 objects.
To investigate whether BPT valley sources have a similar variability feature as shown in our luminosity declining AGN candidates, we also calculate the $\log(R)$ and $\Delta$W1 for the BPT valley population in our sample.
The BPT valley population shows $\log(R)=-0.32\pm 0.40$, which is slightly lower than the value of the parent sample of $\log(R)=-0.02\pm0.37$ and their significance is statistically significant with the $p$-value of $\sim10^{-16}$. 
Our luminosity declining AGN candidates show lower $\log(R)$ of $-1.07\pm 0.09$ and therefore also the p-value
of the distribution difference between the luminosity decline AGN candidates and the BPT valley population is $\sim10^{-16}$. Thus, BPT valley population and luminosity decline AGN is essentially the different population in terms of the obtained $R$ values.
That is, a low-metallicity environment alone cannot reproduce a such significant AGN luminosity decline.

The \textit{WISE} variability is $\Delta W_1=0.39$ for the BPT valley population, which is between the luminosity declining AGN candidates ($\Delta W_1=0.41$) and the parent sample ($\Delta W_1=0.36$).
The BPT valley and the parent sample show a different distribution for the variability this is also suggested by the $p$-value of 0.02, but show a similar one to our luminosity declining AGN candidates with the $p$-value of 0.6.

It is naively expected that the stellar mass of low-metallicity AGN
would reside in a relatively low stellar-mass galaxies, i.e., $M_\star < 10^{10} \msun$, as shown in star-forming galaxies at each cosmic epoch \citep[e.g.,][]{2004Tremonti,2006Lee,2006Erb}.
In addition, some studies suggest that AGN in low stellar-mass galaxies tend to show a higher AGN variability amplitude in the optical bands \citep[e.g.,][]{2020Kimura,2021Burke}, 
while some report that mass dependence is weak \citep{2017Caplar}.
On the other hand, it is not still clear whether 
low-metallicity AGN reside in low-stellar mass galaxies.
Considering that our sample is type-1 AGN, 
the stellar-mass is hard to be constrained with the current sample set.
Instead, we compare the BH mass distributions of each subgroup
as an indicator of the stellar-mass which is inferred from
the scaling relation between $\mbh$ and $\mstar$ \citep[e.g.,][]{2013Kormendy}.
The median $\mbh$ of each subgroup is 
$\left< \log(M_\mathrm{BH}/\msun) \right> = 8.0$, $7.8$, and $7.8$
for luminosity declining AGN candidates, BPT-valley AGN, and the parent sample, respectively.
Assuming the relation between $\mbh$ and $\mstar$ of \cite{2013Kormendy},
the expected stellar-mass is 
$\left< \log(M_\star/\msun) \right> = 10.3$, $10.2$, and $10.2$, respectively.
This suggests that most of luminosity declining AGN candidates do not reside
in low-stellar mass galaxies with $M_\star < 10^{10} \msun$, but 
hey reside in relatively massive ones.
The BPT valley AGN also show similar trend and this relatively
massive stellar-mass of the host galaxies with $M_\star > 10^{10} \msun$
is consistent with the result of \cite{2017Kawasaki}.
One possible mechanism of such low metallicity AGN in the relatively massive host galaxies can be realized if the inflow of the low metallicity gas occurs from
the IGM and/or surrounding environment \citep{2011Husemann}.

\section{Discussion}

\subsection{How long does super-Eddington phase last?}
\label{chap:burst}

Our study shows that there are variable AGN in the time span of $10^{3-4}$~yr, and
their AGN luminosities used to reach around the Eddington limit.
This suggests that the lifetime of such burst phase around the Eddington limit ($t_\mathrm{burst}$) may not last long, a timescale of $10^{3-4}$~yr.
Since the Eddington ratio is estimated for all sample in this study at the three epochs as shown in Figure~\ref{fig:lookback}, we here estimate $t_\mathrm{burst}$ from the number fraction of such super-Eddington phase at each epoch.

We first count the number of sources above the Eddington limit 
at each AGN indicator, $N_{i,\mathrm{Edd}}$ (where $i=$NLR, torus or AD),
and if the sources are above the Eddington limit in the two epoch, those numbers are written
as $N_{i+j,\mathrm{Edd}}$, where $i,j=$NLR, torus, and AD and $D_i > D_j$.
Since the $\lambda_\mathrm{Edd}$ has a certain error, we treat the source as super-Eddington sources only when the $\lambda_\mathrm{Edd} - 3 \Delta \lambda_\mathrm{Edd} > 1$.
Then we calculate the number fraction which are beyond the 
Eddington both in the torus and NLR indicators,
and the obtained fraction is $f_\mathrm{Edd} = N_\mathrm{NLR+torus,Edd}/N_\mathrm{NLR,Edd} = 15 / 114 = 0.13$. We assume that the burst phase lasts with a stable luminosity for the timespan of $t_\mathrm{burst}$. In this case, $f_\mathrm{Edd}$ depends on the $t_\mathrm{burst}$ phase as following

\begin{equation}
f_\mathrm{Edd} =\exp\left( - \dfrac{\Delta t}{t_\mathrm{burst}}\right),
\end{equation}
where $\Delta t$ is the lookback time difference between the torus and NLR, and essentially it is dominated by the look back time to the NLR, with $\Delta t \sim 10^{3.8\pm0.1}$~yr (see Figure~\ref{fig:lookback}). Thus, $t_\mathrm{burst} = -\Delta t / \log(f_\mathrm{Edd}) \simeq 10^{3.5}$~yr 
\footnote{Using a $2\sigma$ instead of $3\sigma$ selection criterion for super Eddington phases would give us $t_\mathrm{burst}\simeq 10^{3.5}$~yr, which has little impact on the estimation of $t_\mathrm{burst}$.}.
This is four-to-five orders of magnitude shorter than the total lifetime of AGN of $\sim10^8$~yr \citep{2004Marconi}, and even one to two orders of magnitude shorter than the typical one-cycle ($t_\mathrm{AGN}$) AGN lifetime of $10^{5}$~yr \citep{2015Schawinski}. 
Therefore, super-Eddington phase can be achieved a certain fraction of the AGN lifetime of $f_\mathrm{burst,Edd} = t_\mathrm{burst}/t_\mathrm{AGN} \sim 0.01$--$0.1$.

Although $f_\mathrm{burst,Edd}$ is small and our sample is based on low-$z$ high-luminosity type-1 AGN in the low-$z$ universe at $z<0.4$,
the suggested $f_\mathrm{burst,Edd}\sim 0.1$ might alleviate 
the current challenge to the quasar evolution is seen in $z>6$, where most of the luminous quasars require the Eddington-limit accretion with the duty cycle of nearly one \citep{2010Willott,2011Mortlock,2015Wu,2018Banados,2021Wang,2021Yang}.
Assuming that the luminous quasars experience a super-Eddington phase with 10\% of their quasar lifetime, and the accretion rate can exceed the Eddington limit by a factor of up to $\sim10$--$100$ \citep{2005Ohsuga,2014Jiang,2015sadowski,2016Inayoshi_b,2020Inayoshi}, which reduces the required quasar growth time by a factor of 2--10.

Recently, the lifetime of high-$z$ ($z>6$) quasars have been reported through the measurements 
of the physical extents of hydrogen Ly$\alpha$ proximity zones \citep{2020Devies,2018Eilers,2021Eilers}.
Some sources show a small physical size, and therefore the inferred quasar lifetime is substantially short, with the order of $10^{3-4}$~yr.
Given that the inferred quasar lifetimes of such $z>6$ quasars are substantially shorter
than the e-folding timescale of the Eddington-limited accretion with $t_\mathrm{Edd} \simeq 4.5\times 10^6$~yr, \cite{2021Inayoshi} suggested that those quasars are expected to experience the super-Eddington accretion phase to grow up. This short timescale of $10^{3-4}$~yr is consistent with our expected $t_\mathrm{burst}$ of the super-Eddingon phase of local AGN at $z<0.4$.
This might be a coincidence, but it is seen in both quasars at $z>6$ and $z<0.4$. If a constant $t_\mathrm{burst}\sim0.1$ of the super-Eddington phase is ubiquitously seen across the cosmic epoch, the lifetime of such super-Eddington phase would be fundamentally governed by the BH accretion disk physics and unrelated to the cosmological environment
once enough gas supply is achieved into the central engine.

\subsection{What are luminosity declining AGN candidates in this study?}
Our goal is to search for luminosity declining AGN who has experienced a large flux decline in the past $10^{3-4}$~yrs, and we selected 57 \othree~bright and MIR faint AGN as luminosity declining AGN candidates.
Figure~\ref{fig:lookback} exhibits
that our method selects sources that experienced burst phase reaching the Eddington limit, and rapidly declined AGN luminosities
at least by a factor of 10 in the last 
$10^{3-4}$~yr.

In addition to such feature in the longterm flux change of $\sim10^{3-4}$~yr span, a
certain fraction of such luminosity declining AGN candidates also show variabilities even in the last $\sim10$~yr scale, which has been 
supported from the \textit{NEOWISE} light curve and the SDSS multi-epoch spectra with high cadence of the changing-look AGN feature. This suggests that our method turn out to be an efficient way to select not only for a long-term AGN variability of $\sim10^{3-4}$~yr, but also for a relatively shorter-term AGN variability with $<10$~yr timescale.

Several authors have already discussed that the flux change of $\sim10$~yr timescale, notably for changing-look AGN, has a different physical origin from the one shown in fading AGN with $\sim10^{3-4}$~yr timescale.
The latter timescale is roughly consistent with the viscous timescale, an inflow timescale of the gas accretion, and can be realized once rapid accretion rate change occurs in the accretion disk \citep[e.g., see discussions in ][]{Kohei19NeoRadio,Kohei19b}.
It was discussed that the origin of the flux change in the timespan of $\sim$yr to $\sim10$~yr rather originates from the instability of the accretion disk \citep{2015LaMassa,2016Ruan,MacLoadCLSearchPhoto,Stern18}.

One key insight obtained from this study is that 
the two epoch spectra are obtained with the timespan of 
1--14~yr, and the average BH mass of the sample is $\left<\mbh \right>\sim 10^8$~$\msun$.
This suggests that the observed changing-look behavior should 
occur within $\sim10$~yr in the system of $\left<\mbh \right>\sim 10^8$~$\msun$. This can rule out some disk timescales
as origins of the changing-look behavior.
For example, the heating and cooling fronts propagation
instability in the disk occurs in the timescale of $t_\mathrm{front} \sim 20 (\mbh/10^8 \msun)$~yr by assuming that the scale
height of the disk is $h/R=0.05$, the disk viscous parameter of $\alpha=0.03$. Since our observing epoch covers the time span of 14~yr at the maximum, the characteristic
cooling front timescale might match with
the changing-look events shown in our sources.
On the other hand, the thermal timescale of the disk,
which can be written as $t_\mathrm{th} \sim 1\times(\mbh/10^8 \msun)$~year,
is slightly too short.

 \cite{2018Noda} discussed such changing-look AGN behavior is tightly connected to the state change and such change preferentially occurs at a specific Eddington ratio range of around $\log \lambda \sim -1$ to $-1.5$ and a timescale of several years.
Considering that most of our luminosity declining AGN candidates have Eddington ratio of around this range and their timescales are also consistent with the expectation from \cite{2018Noda}, this might be related to such state change and our sample might preferentially pick up changing-look AGN with the association of the state change.

In summary, the origins of the possible connection of
the luminosity change between the 10~yr timescale and $10^{3-4}$~yr
are still uncertain, but the preference of the changing-look behavior
at the specific Eddington ratio of $\log \eddington \sim -1$ to $-1.5$ 
and the $\sim10^3$~yr long burst phase at the Eddington ratio of $\log \eddington \sim 0$ suggests that both timescales might connect to the state transition events
in the accretion disk. 
Although the current sample still limits the sample size to only 4~sources who experienced the variabilities both in 10~yr and $10^3$~yr,
the incremental sample will happen soon once the \textit{eROSITA} \citep{pre21} X-ray all-sky data become public at the beginning of 2023.
\textit{eROSITA} will provide the current AGN luminosity for our type-1 AGN sample without worrying about the dust or gas obscuration to the line of sight.
Also, the 0.5--2~keV flux limit of \textit{eROSITA} in the final integration of the planned four years program (eRASS8) in the ecliptic equatorial region \citep[$f_\mathrm{0.5-2keV} = 1.1 \times 10^{-14}~$~erg~s$^{-1}$~cm$^{-2}$][]{pre21}
is way deeper than the expected one estimated from the \textit{WISE} W3 (12~$\mu$m) band flux density limit, where $f_{\nu,\mathrm{lim}} = 1$~mJy corresponds to $f_\mathrm{0.5-2keV} = 8.3 \times 10^{-14}$~erg~s$^{-1}$~cm$^{-2}$, by assuming the local X-ray--MIR luminosity correlation of AGN \citep{gan09,asm15,ich12,ich17a}.

\subsection{Radio Properties of Luminosity Declining AGN}

We here discuss the radio properties of the luminosity declining AGN
since most of the \othree\ luminous extremely red quasars at $z\sim2$--$3$ \citep{ros15,ham17}
are known to be radio bright sources with $\nu L_\mathrm{1.4GHz} \simeq 10^{40-41}$~erg~s$^{-1}$ \citep[e.g.,][]{zak14,hwa18}, which likely originate from shocks caused by wide-angle quasar winds, and
it is worth to check whehter our sources could be local ($z<0.4$) analogous to them.

\cite{MULLANEY} summarized the radio properties of our parent sample
based on the 1.4~GHz radio detections by the VLA/FIRST \citep{bec95,whi97,hel15} 
and NVSS \citep{con98} surveys at a limiting flux density of $>2$~mJy, by following the similar manner of \cite{bes05}.
Out of our 57 luminosity declining AGN candidates, 7 sources have such radio detections
with the median radio luminosity of $L_\mathrm{1.4GHz}\simeq 10^{24}$~W~Hz$^{-1}$ or $\nu L_\mathrm{1.4GHz} \simeq 10^{40}$~erg~s$^{-1}$, whose radio to bolometric luminosity ratio is 
$\log (\nu L_\mathrm{1.4GHz}/\lbolothree) \simeq -6.2$, which is comparable
with the value expected from the quasar wind scenario \citep[e.g.,][]{hwa18}.

The radio detection rate of luminosity declining AGN is $7/57\simeq0.12$, 
which is slightly higher 
than that of our parent sample of $460/7653\simeq 0.06$,
and is closer to the detection rate of the extremely red quasars of $9/97\simeq0.09$
at a same radio survey depth.
Although the sample size is too small at this stage, 
such higher radio detection rate might be a result of the enhancement
of the radio emission by the shocks from the AGN radio driven outflow, 
as discussed in \cite{zak14}.
If our luminosity declining AGN are in the similar population of
extremely red quasar at $z=2$--$3$, luminosity declining AGN 
are also likely in the act of the strong AGN feedback phase with a strong radiative outflow,
and this scenario is also consistent with the experience of the AGN luminosity declining over the past $10^{3-4}$~yr, partially because of the shortage supply of the gas into the nucleus.
However, we also note that the additional deeper radio observations are necessary to
confirm that the similar radio properties can be obtained for the remaining FIRST or NVSS non-detected
luminosity declining AGN.

Some might also wonder that blazars might contaminate the sample of luminosity declining AGN.
For the 7 radio detected sources in our sample, their optical spectra
are dominated by the strong emission lines as well as the blue continuum,
which is a natural outcome based on our selection criteria of type-1 AGN with strong \othree~emission lines with the signal-to-noise ratio of S/N$\geq 5$. This rules out a possibility that they are BL Lac sources whose optical continuum should not show any emission lines.
In addition, their 1.4~GHz radio luminosity is around $\nu L_\mathrm{1.4GHz}\sim10^{40}$~erg~s$^{-1}$.
This is at least one order of magnitude fainter than the typical observed radio luminosity range of the flat spectrum radio quasars \citep[e.g.,][]{ghi17}. Based on those results,
we conclude that blazar contamination is unlikely for our 7 radio detected luminosity declining AGN.

\section{Conclusion}

We systematically search for an AGN population who has experienced
a significant AGN luminosity decline in the past $10^{3-4}$~yr
by utilizing the advantage of the difference of the physical size of 
each AGN indictor, spanning from $<1$~pc to $\gtrsim1$~kpc.
We cross-matched the $\sim7,700$ SDSS DR7 type-1 AGN at $z<0.4$ \citep{mul11}, 
covering the [O~{\sc{iii}}]$\lambda5007$ emission line which is a tracer for the $\sim$kpc scale narrow-line region (NLR) emission, with the \wise~IR catalog which traces the AGN dust emission in the central $\sim10$~pc scale.
With our selection of at least one magnitude fainter in the AGN dust luminosity than the one from the NLR,
we selected an interesting population of the luminosity declining AGN candidates who has experienced the AGN luminosity decrease by a factor of $>10$ in the previous $10^{3-4}$~yr. 
The sample contains 57 AGN and our results show interesting properties which give key insights to the BH and accretion disk physics as written below.

\begin{enumerate}
    \item  The parent type-1 AGN sample shows on average the constant Eddington ratio over the previous $10^{3-4}$~yr, indicating that the intrinsic AGN variability within $\sim10^{3-4}$~yr should be on average smaller than the scatter of 0.4~dex. On the other hand, the luminosity decline AGN candidates show a large luminosity decline in the previous $10^{3-4}$~yr and their previous AGN luminosities reached almost Eddington limit of $\log \eddington \sim 0$, while the current Eddington ratio is almost similar with those of the parent sample of $\eddington \sim -1.5$, indicating the drastic luminosity decline by a factor of $>10$.
    
    \item Utilizing the 3.4~$\mu$m light-curves obtained from \textit{ALLWISE} and \textit{NEOWISE}, the luminosity decline AGN candidates show a relatively larger \textit{W1}-band variability of 0.41 compared to the parent sample of 0.35. In addition, two sources show a continuous flux decline over $\sim10$~yr, suggesting that at least two sources still experience the luminosity decline in the last 10~yr, a possible continuous flux decrease over $10^4$~yr.
    
    \item Thirteen out of the 57 luminosity declining AGN candidates have multi-epoch SDSS spectra and two out of them show a spectral type change associating with a disappearing continuum and broad line emissions, which is so called a changing-look phenomenon. The finding rate is $15$\%, which is two to three orders of magnitude higher than the random selection of the changing-look AGN. Combined with the higher IR varibility signature in the \textit{WISE} bands, our method to select variability AGN in the past $10^{3-4}$~yr might also be an efficient method to select AGN who has recently experienced the AGN variability in the past $\sim10$~yr.
    
    \item The location of the luminosity declining AGN candidates in the BPT-diagram is different from the the parent type-1 AGN sample,
    notably a lower median flux ratio of $\log$([N~{\sc{ii}}]$\lambda6584/ \mathrm{H}\alpha\lambda6563)=-0.52$ than $\log$([N~{\sc{ii}}]$\lambda6584/ \mathrm{H}\alpha\lambda6563)=-0.24$ for the parent sample.
    Considering that the lower flux ratio indicates that their NLR gas has a lower gas metallicity, luminosity declining AGN candidates might prefer the host galaxies with younger and gas rich host galaxies, resulting the past AGN burst reaching the Eddington limit accretion. 
    
    \item Utilizing this long-term light-curve, we estimate the lifetime of the burst phase realizing the super-Eddington accretion. The estimated lifetime is $t_\mathrm{burst}\simeq 10^{3.5}$~yr, suggesting that the super-Eddington phase can be achieved only in a certain fraction of the AGN lifetime of $t_\mathrm{burst}/t_\mathrm{AGN} \sim 0.01$--$0.1$.
    
\end{enumerate}

\acknowledgments

This work is supported by Program for Establishing a Consortium for the Development of Human Resources in Science
and Technology, Japan Science and Technology Agency (JST) and is partially supported by Japan Society for the Promotion of Science (JSPS) KAKENHI (20H01939, K.~Ichikawa; JP20K14529, T.~Kawamruo; 19K21884, 20H01941, and 20H01947, H. Noda).
J.~Pflugradt and K.~Ichikawa also thank Max Planck Institute for hosting us and some of the studies are conducted at MPE.

\software{Astropy \citep{astropy:2018}, Matplotlib \citep{matplotlib}, Pandas \citep{Pandas}}

\bibliographystyle{aasjournal}
\bibliography{bibtex}

\end{document}